\documentclass{emulateapj}
\newcommand{\aeq}{a_{\rm eq}}
\newcommand{\de}{\phi}
\newcommand{\partrack}{{\varepsilon_{\phi \infty}}}
\newcommand{\parslope}{\varepsilon_{s} }
\newcommand{\parrun}{\zeta_{s} }
\newcommand{\omm}{{\Omega_{m0}}}
\newcommand{\omphi}{{\Omega_\phi}}
\newcommand{\phiini}{{\phi_{\rm ini}}}

\newcommand{\mpl}{{M_{\text{p}}}}
\newcommand{\hunit}{{km\;s\textsuperscript{-1}Mpc\textsuperscript{-1}}}
\newcommand{\figurewidth}{3.2in}
\newcommand{\narrowfigurewidth}{2.9in}
\newcommand{\halffigurewidth}{1.7in}
\newcommand{\be}{\begin{equation}}
\newcommand{\ee}{\end{equation}}
\newcommand{\bea}{\begin{eqnarray}}
\newcommand{\eea}{\end{eqnarray}}

\renewcommand{\cite}{\citep[][]}

\def \refeq [#1]{(\ref{#1})}

\usepackage{amsmath,amssymb}

\begin{document}
\title{Parameterizing and Measuring Dark Energy Trajectories from Late-Inflatons}
\author{Zhiqi~Huang\altaffilmark{1,2}, J.~Richard~Bond\altaffilmark{1} and Lev~Kofman\altaffilmark{1}}
\altaffiltext{1}{Canadian Institute for Theoretical Astrophysics, University of Toronto, Toronto, Canada}
\altaffiltext{2}{Department of Astronomy \& Astrophysics, University of Toronto, Toronto, Canada}
\begin{abstract}
Bulk dark energy properties are determined by the redshift evolution of its pressure-to-density ratio, $w_{\rm de}(z)$. An experimental goal is to decide if the dark energy is dynamical,  as in the quintessence (and phantom)  models treated here. We show that a three-parameter approximation $w_{\rm de}(z; \parslope, \partrack, \parrun)$ fits well the ensemble of trajectories for a wide class of  late-inflaton potentials $V(\phi)$. Markov Chain Monte Carlo probability calculations are used to confront our $w_{\rm de}(z)$ trajectories with current observational information on Type Ia supernova, Cosmic Microwave Background, galaxy power spectra, weak lensing and the Lyman-${\alpha}$ forest. We find the best constrained parameter is  a low redshift slope parameter, $\parslope \propto (\partial \ln V / \partial \phi)^2$ when the dark energy and matter have equal energy densities. A tracking parameter $\partrack$ defining the high-redshift attractor of $1+w_{\rm de}$ is marginally constrained. Poorly determined is $\parrun$, characterizing the evolution of $\parslope$, and a measure of $\partial^2 \ln V / \partial \phi^2$ . The constraints we find already rule out some popular quintessence and phantom models, or restrict their potential parameters. We also forecast how the next generation of cosmological observations improve the constraints:  by a factor of about five on $\parslope$ and $\partrack$, but with $\parrun$ remaining unconstrained (unless the true model significantly deviates from $\Lambda$CDM). Thus potential reconstruction beyond an overall height and a gradient is not feasible for the large space of late-inflaton models considered here. 
\end{abstract}

\setcounter{footnote}{0}
\section{Introduction}\label{sec:intro}

\subsection{Running Dark Energy and its Equation of State}

One of the greatest mysteries in physics is the nature of dark energy (DE) which drives the present-day cosmic acceleration,  inferred to exist from supernovae data \citep{Riess1998,Perlmutter1999}, and from a combination of cosmic microwave background and large scale structure data \citep{Bond1999}. Although there have been voluminous outpourings on possible theoretical explanations of dark energy, recently reviewed in \citet{Copeland2006, Kamionkowski2007,Linder2008b,Linder2008a,Silvestri2009}, we are far from consensus. An observational target is to determine if there are temporal (and spatial) variations beyond the simple constant $\Lambda$. Limits from the evolving data continue to roughly center on the cosmological constant case $\Lambda$, which could be significantly strengthened in near-future experiments -- or ruled out. We explore the class of effective scalar field models for dark energy evolution in this paper, develop a 3-parameter expression which accurately approximates the dynamical histories in most of those models, and determine current constraints and forecast future ones on these parameters. 

The mean dark energy density is a fraction $ \Omega_{\rm de}$ of the mean total energy density, 
\begin{equation}\label{eq:Omegade}
\rho_{\rm de} (\ln a)\equiv \rho_{\rm tot}  \Omega_{\rm de}\, , \ 3\mpl ^2 H^2(a)= \rho_{\rm tot}(a)\, , 
\end{equation}
which is itself related to the Hubble parameter $H$ through the energy constraint equation of gravity theory as indicated. $\Omega_{\rm de}(a)$  rises from a small fraction relative to the matter at high redshift to its  current $\sim 0.7$ value. Here $\mpl =1/\sqrt{8\pi G_N} =2.44\times10^{18}\text{GeV}$,   is the reduced Planck mass, $G_N$ is Newton's gravitational constant and $\hbar$ and $c$ are set to unity, as they are throughout this paper.  

Much observational effort is being unleashed to determine as much as we can about the change of the trajectory $\rho_{\rm de} $ with expansion factor $a$, expressed through a logarithmic running with respect to $\ln a$: 
\begin{equation}\label{eq:Pidot}
  -{1 \over 2} {d\ln \rho_{\rm de} \over d\ln a} \equiv \epsilon_{\rm de}(\ln a) \equiv {3\over 2}(1+w_{\rm de}) \equiv {3\over 2}{(\rho_{\rm de}+p_{\rm de})\over \rho_{\rm de}} . 
\end{equation}
This $\rho_{\rm de}$-run is interpreted as defining a phenomenological average pressure-to-density ratio, the dark energy equation of state (EOS). The total ``acceleration factor'',  $\epsilon \equiv 1+q$, where the conventional ``deceleration parameter'' is $q\equiv -a\ddot{a}/\dot{a}^2$ = $-d\ln (Ha) /d\ln a$, is  similarly related to the running of the total energy density:  
\begin{equation}\label{eq:acc}
 -{1 \over 2}  {d\ln \rho_{\rm tot} \over d\ln a} \equiv \epsilon = \epsilon_{m}\Omega_m + \epsilon_{\rm de}\Omega_{\rm de} \, , \  \Omega_m + \Omega_{\rm de}=1\, .  
\end{equation}
Eq.~\refeq[eq:acc] shows $\epsilon $ is the density-weighted sum of the acceleration factors for matter and DE. Matter here is everything but the dark energy. Its EOS has  $\epsilon_{m}=3/2$ in the non-relativistic-matter-dominated phase (dark matter and baryons) and $\epsilon_{m}=2$ in the radiation-dominated phase.

\subsection{The Semi-Blind Trajectory Approach}

We would like to use the data to constrain $\rho_{\rm de}(\ln a) $ with as few prior assumptions on the nature of the trajectories as is feasible. However, such blind analyses are actually never truly blind, since $\rho_{\rm de}$ or $w_{\rm de}$ is expanded in a truncated basis of mode functions of one sort or another: necessarily there will  be assumptions made on the smoothness  associated with the order of the mode expansion and on the measure,  i.e., prior probability,  of the unknown coefficients multiplying the modes. The  most relevant current data for constraining $w_{\rm de}$, SNIa compilations, extends back only about one e-folding in $a$, and probes a double integral of $w_{\rm de}$, smoothing over irregularities. The consequence is that unless $w_{\rm de}$ was very wildly varying, only a few parameters are likely to be extractable no matter what expansion is made. 

\begin{figure}
\centering
\includegraphics[width=\figurewidth]{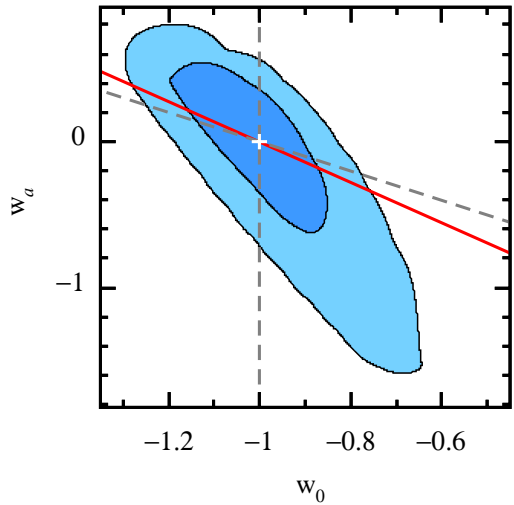}
\caption{The marginalized 68.3\% CL (inner contour) and 95.4\% CL (outer contour) constraints on $w_0$ and $w_a$ for the conventional DETF parametrization $w=w_0+w_a(1-a)$, using the current data sets described in \S~\ref{sec:constraint}. The white point is the cosmological constant model. The solid red line is a slow-roll consistency relation, Eq.~\refeq[slowroll_constr] derived in \S~\ref{sec:w0wadegeneracy} (for a fixed $\Omega_{m0}=0.29$, as inferred by all of the current data). The tilted dashed gray line shows $w_a =-1-w_0$. Pure quintessence models restrict the parameter space to $1+w_0 > 0 $ and $w_a$ above the line, whereas in the pure phantom regime,  $1+w_0 <-0$ and $w_a$ would have to lie below the line. Allowing for equal a priori probabilities to populate the other regions is quite problematical theoretically, and indeed the most reasonable theory prior would allow only the pure quintessence domain. }\label{figw0wa2D}
\end{figure}

Low order expansions include the oft-used but nonphysical cases of constant $w_0 \ne -1$ and the 2-parameter linear expansion in $a$  \citep{Chevallier2001,Linder2003,Wang2004,Upadhye2005,Liddle2006,Francis2007,Wang2007},    
\begin{equation}\label{conv}
w(a)=w_0+w_a (1-a) \ , 
\end{equation}
adopted by the Dark Energy Task Force (DETF) \citep[see][]{Albrecht2006}. 
The current observational data we use in this paper to constrain our more physically-motivated parameterized trajectories are applied in Figure~\ref{figw0wa2D} to $w_0$ and $w_a$, assuming uniform uncorrelated priors on each. The area of the nearly-elliptical 1-sigma error contour has been  used to compare how current and proposed dark energy probes do relative to each other; its inverse defines the DETF ``figure of merit'' (FOM). 

Constant $w_{\rm de}$ and Eq.~\refeq[conv] can be considered to be the zeroth and first order polynomial expansion in the variable $1-a$.  Why not redshift $z$ \cite{Upadhye2005, Linder2005} or the scale factor to some power \cite{Linder2005} or $\ln a$? Why pivot about $a=1$. 

Why only linear in $a$? Why not expand to higher order? Why not use localized spline mode functions, the simplest of which is a set of contiguous top hats (unity in a redshift band, zero outside)? These cases too have been much explored in the literature, both for current and forecast data. The method of principal component (parameter eigenmode) analysis takes orthogonal linear combinations of the mode functions and of their coefficients, rank-ordered by the error matrix eigenmodes \citep[see e.g.,][]{Huterer2003,Crittenden2005}.  As expected only a few linear combinations of mode function coefficients are reasonably determined.  Thus another few-parameter approach expands $w_{\rm de}$ in many modes localized in redshift but uses only the lowest error linear combinations. At least at the linearized level, the few new parameters so introduced are uncorrelated. However, the eigenmodes are sensitive to the data chosen and the prior probabilities of the mode parameters. An alternative to the $w_{\rm de}$ expansion is a direct mode expansion in $\epsilon$, considered at low order by \citet{Rapetti2007}; since the  $\epsilon_{m}$ part is known, this is an expansion in $(\epsilon_{\rm de}-\epsilon_{m})\Omega_{\rm de}$,  in which case $w_{\rm de}$  becomes a derived trajectory;  for  $w_{\rm de}$ parameterizations it is  $\epsilon (a)$ that is the derived trajectory. 

The partially blind expansions of  $1+w_{\rm de}$ are similar to the early universe inflation expansions of the scalar power spectrum ${\cal  P}_s (\ln k)$, e.g., \citet{scanning}, except in that case one is ``running'' in ``resolution'',  $\ln k$. However there is an approximate relation between the comoving wavenumber $k$ and the time when that specific $k$-wave went out of causal contact by  crossing the instantaneous comoving ``horizon''  parameterized by $ Ha$.  This allows one to translate the power spectra $\ln k$-trajectories into dynamical $\ln Ha$-trajectories, ${\cal  P}_s (\ln Ha)$. Lowest order is a uniform slope $n_s -1 = d\ln {\cal P}_s /\ln k$; next order is a running of that slope, rather like the DETF linear $w_{\rm de}$ expansion. Beyond that, one can  go to higher order, e.g., by expanding in Chebyshev polynomials in $\ln k$, or by using localized modes to determine the power spectrum in $k$-bands. Generally there may be tensor and isocurvature power spectra contributing to the signals, and these would have their own expansions. 

In inflation theory, the tensor and scalar spectra are  related to each other by an approximate consistency condition:  both are derivable from parameterized acceleration factor trajectories defining the inflaton equation of state, $1+w_{\phi}(a) = 2\epsilon /3$ (with zero $\Omega_m$ in early universe inflation). Thus there is a very close analogy between phenomenological treatments of  early and late universe inflation. However, there is a big difference: power spectra can be determined by over $\sim$10 e-foldings in $k$ from CMB and Large Scale Structure data on clustering, hence over $\sim$10 e-foldings in $a$, whereas dark energy data probe little more than an e-folding of $a$.  This means that higher order partially blind mode expansions are less likely to bear fruit in the late-inflaton case. 

\subsection{Physically-motivated Late-inflaton Trajectories}

This arbitrariness in semi-blind expansions motivates our quest to find a theoretically-motivated prior probability on general dark energy trajectories characterized by a few parameters, with the smoothness  that is imposed following from physical models  of their origin rather than from arbitrary restrictions on blind paths. Such a prior has not been much emphasized in dark energy physics, except on an individual-model basis. The theory space of models whose trajectories we might wish to range over include:

\noindent
(1) quintessence, with the acceleration driven by a scalar field minimally coupled to gravity with an effective potential with small effective mass and a canonical kinetic energy  \cite{Ratra1988,Wetterich1988,Frieman1995,Binetruy1999,Barreiro1999,Brax1999,Copeland2000,Macorra2001,Linder2006,Huterer2007,Linder2008c}; 

\noindent
(2) k-essence, late inflation with a non-canonical scalar field kinetic energy as well as a potential  \cite{Armendariz2001,Malquarti2003a,Malquarti2003b,Putter2007}; 

\noindent
(3) $F(R,\phi)$ models, with Lagrangians involving a function of the curvature scalar $R$ or a generalized dilaton $\phi$, that can be transformed through a conformal mapping into a theory with an effective scalar field, albeit with interesting fifth force couplings to the matter sector popping out in the transform  \cite{Capozziello2002,Carroll2004,deLaCruz-Dombriz2006,Nojiri2006,deFelice2010,Sptiriou2010}; 

\noindent
(4) phantom energy with $w_{\rm de} < -1$, and negative effective kinetic energy, also with a potential \cite{Copeland2006,Carroll2003,Caldwell2002,Caldwell2003} . 

In this paper,  we concentrate on the quintessence class of models, but allow trajectories that would arise from a huge range of potentials, with our fitting formula only deviating significantly when $|1+w_{\rm de}| > 0.5$. A class of models that we do not try to include are ones with dark energy having damped oscillations before settling into a minimum $\Lambda$ \cite{Kaloper2006,Dutta2008}. 

We also extend our paths to include ones with $w_{\rm de} < -1$, the phantom regime where the null energy condition \cite{Carroll2003} is violated. This can be done by making the kinetic energy negative, at least over some range, but such models are ill-defined. There have been models proposed utilizing extra UV physics, such as with a Lorentz-violating cut-off \cite{Cline2004} or with extra fields \cite{Sergienko2008}.  Regions with $w_{\rm de} <-1$ are consistent with current observations, so a semi-blind phenomenology should embrace that possibility. It is straightforward to extend our quintessence fitting formula to the phantom region, so we do, but show results with and without  imposing the theoretical prior that these are highly unlikely.

The evolution of $w_{\rm de}$ for a quintessence field $\phi$ depends on its potential $V(\phi)$ and its initial conditions. For a flat Friedmann-Robertson-Walker (FRW) universe with given present matter density $\rho_{m0}$ and dark energy density $\rho_{\rm de 0}$, any function $w_{\rm de}(a)$ satisfying $0\le 1+w_{\rm de}\le 2$ can be reverse-engineered into a potential $V(\phi)$ and an initial field momentum, $\Pi_{\phi, \rm ini} \equiv \dot\phi_{\rm ini}$. (The initial field value $\phi_{\rm ini}$ can be eliminated by a translation in $V(\phi)$, and the sign of $\Pi_{\phi,\rm ini}$ by a reflection in $V(\phi)$.) Given this  one-to-one mapping between $\{ V(\phi), |\dot\phi_{\rm ini}|; \rho_{m0}\}$ and $\{ w(a), \rho_{\rm de 0}; \rho_{m0}\}$, we may think we should allow for all $w_{\rm de}$ trajectories. Generally, these will lead to very baroque potentials indeed, and unrealistic initial momenta. The quintessence models with specific potentials that have been proposed in the literature are characterized by a few parameters, and have a long period in which trajectories relax to attractor ones, leading to smooth trajectories in the observable range at low redshift. These are the simple class of potentials we are interested in here: our philosophy is to consider ``simple potentials'' rather than simple mathematical forms of $w_{\rm de}(a)$. The potentials which we quantitatively discuss in later sections include: power-laws \cite{Ratra1988, Steinhardt1999};  exponentials \cite{Ratra1988}; double exponentials \cite{Barreiro1999,Sen2001,Neupane2003,Neupane2004,Jarv2004};  cosines from pseudo-Nambu Goldstone bosons (pnGB) \cite{Frieman1995,Kaloper2006}; supergravity (SUGRA) motivated models \cite{Brax1999}. Our general method applies to a much broader class than these though: we find the specific global form of the potential is not relevant because the relatively small motions on its surface in the observable range allow only minimal local shape characteristics to be determined, and these we can encode with the three key physical variables parameterizing our $w_{\rm de}$. Even with the forecasts data to come, we can refine our determinations but not gain significantly more information -- unless the entire framework is shown to be at variance with the data.  

\subsection{Tracking and Thawing Models}

Quintessence models are often classified into tracking and thawing models \cite{Caldwell2005,Linder2006,Linder2008c,Huterer2007,Sullivan2007,Scherrer2008}. Tracking models were first proposed to solve the coincidence problem, i.e., why the dark energy starts to dominate not very long after the epoch of matter-radiation equality ($\sim 7.5$ e-folds). However, a simple negative power-law tracking model that solves the coincidence problem predicts $w_{\rm de} \gtrsim -0.5$ today, which is strongly disfavoured by current observational data. In order to achieve the observationally favoured $w_{\rm de}\approx -1$ today, one has to assume the potential changes at an energy scale close to the energy density at matter-radiation equality. The coincidence problem is hence converted to a fine-tuning problem. In this paper we take the tracking models as the high-redshift limit, and parameterize it with a ``tracking parameter'' $\partrack$ that characterizes the attractor solution. The thawing models, where the scalar field is frozen at high redshift due to large Hubble friction, can be regarded as a special case of tracking models where the tracking parameter is zero. Because of the nature of tracking behaviour, where all solutions regardless of initial conditions approach an attractor, we do not need an extra parameter to parameterize the initial field momentum. 

In either tracking models or thawing models the scalar field has to be slowly-rolling or moderately-rolling at low redshift, so that the late-universe acceleration can be achieved. The main physical quantity that affects the dynamics of quintessence in the late-universe accelerating epoch is the slope of the potential. The field momentum is ``damped'' by Hubble friction, and the slope of the potential will determine how fast the field will be rolling down; actually, as we show here,  it is  the $\phi$-gradient of $\ln V$ that matters, and this defines our second ``slope parameter''  $\parslope$.

It turns out that a two-parameter formula utilizing just  $\partrack$ and $\parslope$ works very well, because the field rolls slowly at low redshift, indeed is almost frozen, so the trajectory does not explore changes in the slope of $\ln V$. However, we want to extend our formula to cover the space of moderate-roll as well as slow-roll paths. Moreover, in cases in which $\ln V$ changes significantly, even slow-roll paths may explore the curvature of $\ln V$ as well as its $\phi$-gradient. Accordingly, we expanded our formula to encompass such cases, introducing a third parameter $\parrun$, which is related to the second $\phi$-derivative of $\ln V$, explicitly so in thawing models.

\subsection{Parameter Priors for Tracking and Thawing Models}

Of course, there is one more important fourth parameter, characterizing the late-inflaton energy scale, such as the current dark energy density $\rho_{\rm de, 0}$ at redshift 0. This is related to the current Hubble parameter $H_0\equiv 100h\ \text{km s}^{-1}\text{Mpc}^{-1}$, the present-day fractional matter density $\Omega_{m0}$, and the current fractional dark energy density $\Omega_{\Lambda}\equiv \Omega_{\rm de,0} = 1-\Omega_{m0}h^2/h^2$:  $\rho_{\rm de, 0} =3\mpl ^2 H_0^2 \Omega_\Lambda$. With CMB data, the sound-crossing angular scale at recombination is actually used, as is the physical density parameter of the matter, $\Omega_{m0}h^2$, which is  typically derived from separately determined dark matter and baryonic physical density parameters,  $\Omega_{\rm dm 0}h^2$ and $\Omega_{b0}h^2$. With any nontrivial $w_{\rm de}$ EOS, this depends upon the parameters of $w_{\rm de}$ as well as on one of $h$ or the inter-related zero redshift density parameters. The important point here is that the prior measure is uniform in the sound-crossing angular scale at recombination and in $\Omega_{\rm dm 0}h^2$ and $\Omega_{b0}h^2$, not in $h$ or the energy scale of late-inflation. If the quantities are well-measured, as they are, the specific prior does not matter very much. 

What to use for the prior probability of the parameters of $w_{\rm de}$, in which most are not well-determined? The prior chosen does matter, and has to be understood when assessing the meaning of the derived constraints. Usually a uniform prior in $w_0$ and $w_a$ is used for the DETF form, Eq.~\refeq[conv]. The analogue for our parameterization is to be uniform in the relatively well-determined $\parslope$, but we show what variations in the measure on it do, {e.g.}, using  $\sqrt{|\parslope|}$ or disallowing phantom trajectories,  $\parslope > 0$. The measure for each of the other two parameters is also taken to be uniform. (Our physics-based parameterization leads to an approximate ``consistency relation'' between $w_0$ and $w_a$, a strong prior relative to the usual uniform one.)

In \S~\ref{sec:solution},  we manipulate the dynamic equations for quintessence and phantom fields to derive our approximate solution,  $w_{\rm de}(a\vert \parslope, \partrack, \parrun,\omm)$. We show how well this formula works in reproducing full trajectories computed for a variety of potentials.  In \S~\ref{sec:constraint} we describe our extension of  the CosmoMC program package \cite{Lewis2002} to treat our parameterized DE and the following updated  cosmological data sets: Type Ia Supernova (SN), galaxy power spectra that probe large scale structure (LSS), weak lensing (WL),  Cosmic Microwave Background (CMB), and Lyman-$\alpha$ forest (Ly$\alpha$). The present-day data are used to constrain our dynamical $w_{\rm de}(a)$. In \S~\ref{sec:forecast},  we investigate how future observational cosmology surveys  can sharpen the  constraints on $w_{\rm de}(a\vert \parslope, \partrack, \parrun, \omm)$. We restrict our attention to flat universes. We discuss our results in \S~\ref{sec:discussion}.

\section{Late-inflation Trajectories and Their Parameterization}\label{sec:solution}

\subsection{The Field Equations in Terms of Equations of State}

We assume the dark energy is the energy density of a  quintessence (or phantom) field,  with a canonical kinetic energy and an effective potential  energy $V(\phi)$. This is derived from a Lagrangian density 
\be
{\cal L}=\pm \frac{1}{2}\partial_\mu\phi\partial^\mu\phi-V(\phi) \ , \label{lagrangian}
\ee
where $+$  is for quintessence ($-$ for phantom). Throughout this paper we adopt the $(+,-,-,-)$ metric signature. For homogeneous fields  the energy density,  pressure and equation of state are 
\bea
&& \rho_\phi=\frac{1}{2}\Pi_\phi^2+V(\phi), \ p_\phi=\frac{1}{2}\Pi_\phi^2-V(\phi) ,  \label{pressure} \\
&& 1+w_\phi \equiv \frac{\rho_\phi+p_\phi  }{\rho_\phi} = \frac{\Pi_\phi^2}{\rho_\phi},\ \Pi_\phi\equiv {d\phi \over dt}. \label{wphidef} 
\eea
Since we identify the dark energy with an inflaton, we hereafter use $\phi$ as the subscript rather than ``de''. The field $\phi$ does not have to be a fundamental scalar, it could be an effective field, an order parameter associated with some collective combination of fields. A simple way to include  the phantom field case is to change the kinetic energy sign to minus, although thereby making it a ghost field with unpalatable properties even if it is the most straightforward way to get $1+w_{\rm de} < 0$. 

The two scalar field equations, for the evolution of the field, $\dot{\phi}={\Pi}_\phi$,  and of the field momentum, $\dot{\Pi}_\phi = -3H{\Pi}_\phi-\partial V/\partial \phi$,  transform to 
\begin{eqnarray}\label{eq:epsdot}
\dot{\phi}/H: &&\ \sqrt{\epsilon_{\de}\Omega_{\de}}  = {\pm} d\psi /d\ln a \,  , \  \psi \equiv \phi /( \sqrt{2}\mpl ) \, , \\
\dot{\Pi}_\phi/H: &&\ \frac{d^2\psi}{ (d\ln a)^2}  + 3\left(1-\frac{\epsilon}{3}\right)\, \frac{d\psi}{ d\ln a}   \nonumber \\
&& = \pm 3\sqrt{\epsilon_V \Omega_{\de}} \sqrt{\Omega_{\de}}\left(1-\frac{\epsilon_{\de}}{3} \right)\, .  
\end{eqnarray}
The sign is arbitrary, depending upon whether the field rolls down the potential to larger $\psi$ ($+$) or smaller $\psi$ ($-$) as the universe expands. For definiteness we take it to be positive. 
The second equation can be recast into a first order equation for $d\ln \sqrt{\epsilon_{\de}\Omega_{\de}}/d\ln a$ which is implicitly used in what follows, we prefer to work instead with a first integral, the energy conservation equation~\refeq[eq:Pidot]. 

As long as $\Pi_\phi$ is strictly positive, $\psi$ is as viable a time variable as $\ln a$, although it only changes by $\sqrt{\epsilon_{\de}\Omega_{\de}} $ in an e-folding of $a$. Thus, along with trajectories in $\ln a$, we could reconstruct the late-inflaton potential $V(\phi) $ as a function of $\phi$ and the energy density as a function of the field: 
\begin{eqnarray}\label{eq:epsV}
\sqrt{\epsilon_V} &&\equiv  -{1 \over 2}  {d\ln V\over d\psi} = {\sqrt{\epsilon_{\de}} \over \sqrt{\Omega_{\de}} } \left(1+{{d\ln \epsilon_{\de} /d\ln a} \over {6(1-\epsilon_{\de} /3 ) }}\right)\, ,   \label{epsvdef} \\
\sqrt{\epsilon_E} &&\equiv -{1 \over 2}  {d\ln \rho_{\de}\over d\psi} = {\sqrt{\epsilon_{\de}} \over \sqrt{\Omega_{\de}} } \, . \label{epsedef} \\
\end{eqnarray}
These ``constraint'' equations require knowledge of both $\epsilon_{\de}$ and $\Omega_{\de}$, but the latter runs according to
\be
\frac{1}{2}\frac{d\ln [\Omega_{\de}^{-1}-1]}{d\ln a} = \epsilon_{\de} - \epsilon_m \, , \label{odeo} 
\ee 
hence is functionally determined. Eq.~\refeq[odeo] is obtained by taking the difference  of Eqs.~\refeq[eq:Pidot] and \refeq[eq:acc] and grouping all $\Omega_{\de}$ terms on the left hand side.  

For early universe inflation with a single inflaton,  $\Omega_{\de}=1$, and $\epsilon_{\de} =  \epsilon_E = \epsilon$. The field momentum therefore obeys the relation
 $\Pi_\phi = -2 \mpl ^2 \partial H /\partial \phi$, which is derived more generally from the momentum constraint equation of general relativity \cite{Salopek1990}. 
 
\subsection{A Re-expressed Equation Hierarchy Conducive to Approximation}

The field equations form a complete system if $V(\psi )$ is known, but what we wish to do is to learn about $V$. So we follow the running of a different grouping of variables, namely the differential equations for the set of parameters $\sqrt{\epsilon_{\de}}$  and $\sqrt{\epsilon_{V}\Omega_{\de}}$, with a third equation for the  running of $\Omega_{\de}$, Eq.~\refeq[odeo]:
\bea
\frac{d\sqrt{\epsilon_{\de}}}{d\ln a} &=& 3 (\sqrt{\epsilon_V\Omega_{\de}} -\sqrt{\epsilon_{\de}} )  (1-\epsilon_{\de}/3) \ , \label{odex} \\
\frac{d\ln \sqrt{\epsilon_V\Omega_{\de}}}{d\ln a} &=&   (\epsilon_{\de}-\epsilon_m)(1-\Omega_{\de}) + 2\gamma \sqrt{\epsilon_{\de}} \sqrt{\epsilon_V \Omega_{\de}} .  \label{odel}
\eea
The reason for choosing Eq.~\refeq[odel] for $\sqrt{\epsilon_V\Omega_{\de}}$ rather than the simpler running equation in $\sqrt{\epsilon_V}$ 
\be
\frac{1}{2}d\ln\sqrt{\epsilon_V}/d\ln a = \gamma  \sqrt{\epsilon_{\de}} \sqrt{\epsilon_V \Omega_{\de}} \, ,  \label{odeev}
\ee
 is because we must allow for the possibility that the potential could be steep in the early universe, so 
 $\epsilon_V$ may be large, even though the allowed steepness now is constrained. For tracking models where a high redshift attractor with constant $w_\phi$ exists, $\epsilon_V\Omega_{\de}$ tends to a constant, and is nicely bounded, allowing better approximations. 
 
 (The $\phi \rightarrow -\phi$ symmetry allows us to fix the $\pm \sqrt{\epsilon_{\de}}$ and $\pm  \sqrt{\epsilon_V\Omega_{\de}}$ ambiguity to the positive sign. For phantom energy, we use the negative sign, in Eqs.~(\ref{odex} -\ref{odeev}). 
In these equations, we have restricted ourselves to fields that  always roll down for quintessence, or  up for phantom. The interesting class of oscillating quintessence models \citep{Kaloper2006,Dutta2008} are thus not considered.)
 
These equations are not closed, but depend upon a potential shape parameter $\gamma$, defined by
\begin{equation}
\gamma \equiv\frac{ \partial^2 \ln V /\partial \psi^2 }{ (\partial \ln V /\partial \ln \psi)^2}  \, . \label{gammadef}
\end{equation}
It is related to the effective mass-squared in $H^2$ units through 
\begin{equation}
m_{\rm \de,eff}^2 /H^2 = 6 (1+\gamma)\epsilon_V \Omega_{\de} (1-\epsilon_{\de}/3) \, , 
\end{equation}
Determining how $\gamma$  evolves involves yet higher derivatives of $\ln V$, ultimately an infinite hierarchy unless the hierarchy is closed by specific forms of $V$. However, although not closed in $\sqrt{\epsilon_V\Omega_{\de}}$, the equations are conducive for finding an accurate approximate solution, which we express in terms of a new time variable, 
\begin{equation}
y\equiv \sqrt{\Omega_{\rm \de app}} \, , \ \Omega_{\rm \de app} \equiv \frac{\rho_{\rm \de eq}}{\rho_m(a)+\rho_{\rm \de eq}}= \frac{(a/\aeq )^3}{1+ (a/\aeq )^3} \, , 
\end{equation}
where the subscript ``eq'' defines variables at the redshift of matter-DE equality:
\begin{equation}
\aeq \equiv (\rho_{m0}/\rho_{\rm m,eq})^{1/3}  , \,   \rho_{\rm m,eq}=\rho_{\rm \de eq}, \, \Omega_{\rm \de eq}=\Omega_{\rm m, eq}=1/2 \, .
\end{equation}
Thus $y$ is an approximation to $\sqrt{\Omega_{\de}}$, pivoting about the expansion factor $\aeq $ in the small $\epsilon_{\de}$ limit. By definition $y_{\rm eq}\equiv y\vert_{a=\aeq} = 1/\sqrt{2}$.

\subsection{The Parameterized Linear and Quadratic Approximations}

We now show how working with these running variables leads to a 1-parameter fitting formula, expressed in terms of  
\be
\varepsilon_s \equiv \pm \epsilon_V\vert_{a=\aeq}\, .   \label{parslopedef}
\ee
The 2-parameter fitting formula adds the asymptotic EOS factor
\be
 \partrack  \equiv \pm \epsilon_V\Omega_{\de}\vert_\infty \, .  \label{partrackdef}
\ee
The minus sign in Eq.~\refeq[parslopedef] and \refeq[partrackdef] is a convenient way to extend the parameterization to cover the phantom case.
Here $\infty$ refers to the $a\ll 1$ limit \footnote{For tracking models at high redshift, the asymptotic $\partrack$ is $ \propto \epsilon_m$, and so  varies by a factor of $3/4$ from the radiation-dominated era to the matter-dominated era. In practice we actually use $|\partrack|/\epsilon_m$ as the parameter in our MCMC calculations, hence $\frac{|\partrack|}{\epsilon_m}\epsilon_m{\rm sgn\,}(\parslope)$ replaces $\partrack$ in our $w_\phi$ parametrization.}. The asymptotic equality of our 2 key variables is derived, not imposed, a consequence of the attractor. In addition, if the initial field momentum is far from the attractor, it would add another variable, but the damping to the attractor goes like $a^{-6}$ in $\epsilon_\phi$, hence should be well established before we get to the observable regime for trajectories. 

The 3-parameter form involves, in addition to these two,  a parameter which is a curious relative finite difference of $d\sqrt{\epsilon_V\Omega_{\de}}/dy$ about $y_{\rm eq}/2$:
\be
\parrun \equiv \frac{d\sqrt{\epsilon_V\Omega_{\de}}/dy\vert_{\rm eq} - d\sqrt{\epsilon_V\Omega_{\de}}/dy\vert_{\infty}}{d\sqrt{\epsilon_V\Omega_{\de}}/dy\vert_{\rm eq} + d\sqrt{\epsilon_V\Omega_{\de}}/dy\vert_{\infty}} \ . \label{parrundef}
\ee
The physical content of the ``running parameter'' $\parrun$ is more complicated than that of $\parslope$ and $\partrack$. It is related not only to the second derivative of $\ln V$, through $\gamma$, thus extending the slope parameter $\parslope$ to another order, but also to the field momentum. As we mentioned in the introduction, the small stretch of the potential surface over which the late-inflaton moves in the observable range make it difficult to determine the second derivative of $\ln V$ from the data (\S~\ref{sec:discussion}. In thawing models, for which the field momentum locally traces the slope of $\ln V$, the dependence of $\parrun$ on the field momentum can be eliminated. However,  if the field momentum is sufficiently small, $w_\de $ does not respond to $d^2\ln V/d\phi^2$. We discuss these cases in \S~\ref{sec:discussion}.

What we demonstrate is that a relation linear in $y$, 
\be
  \sqrt{\epsilon_V\Omega_{\de}} \approx \sqrt{\varepsilon_{\de \infty}} + ( \sqrt{\varepsilon_{s}} -  \sqrt{2\varepsilon_{\de \infty}}) y \, , \label{lambdao1}
  \ee
is a suitable approximation. It  yields the 2-parameter formula for $\epsilon_{\de}$, which maintains the basic  form linear in parameters, 
\be
\sqrt{\epsilon_{\de}}  
\approx \sqrt{\varepsilon_{\de \infty}} + \left( \sqrt{\parslope } - \sqrt{2\varepsilon_{\de \infty}} \,  \right )F\left({a\over \aeq}\right) \, ,  \label{eq:2paramsqeps} 
\ee
with
\be
F(x) \equiv  \frac{\sqrt{1+x^3}}{x^{3/2}}-\frac{\ln\left[x^{3/2}+\sqrt{1+x^3}\right]}{x^3}. \label{Fxdef} 
\ee

The 1-parameter case has $\sqrt{\varepsilon_{\de \infty}}$ set to zero, hence the simple $\sqrt{\epsilon_V\Omega_{\de}} \approx  \sqrt{\varepsilon_s\Omega_{\rm \de app}}$ and $\epsilon_\phi = \parslope F^2(a/\aeq )$, which we regard as the logical physically-motivated improvement to the conventional single-$w_0$ parameterization, $1+w= (1+w_{ 0} ) F^2 (a/\aeq )/F^2(1/\aeq )$. 

The approximation for the 3-parameter formula adds a quadratic correction in $y$:
\be
  \sqrt{\epsilon_V\Omega_{\de}} \approx  \sqrt{\varepsilon_{\de \infty}} + (\sqrt{\varepsilon_{s}} -  \sqrt{2\varepsilon_{\de \infty}})y 
 \left[1-  \zeta_s \left(1- {y \over y_{\rm eq}}\right)\right] ,   \label{3paramepsV}
  \ee
and a more complex form for the DE trajectories, 
\be
\sqrt{\epsilon_{\de}} =  \sqrt{\partrack}  + ( \sqrt{\parslope}- \sqrt{2\partrack})\left[ F\left({a\over \aeq}\right) + \zeta_s F_2\left({a\over \aeq}\right)\right] \, , \label{3parameps}
\ee
with
\be
F_2 (x) \equiv \sqrt{2}\left[1-{\ln \left(1+x^3\right) \over x^3}\right] - F(x) \, . \label{F2def}
\ee
To cover the phantom case, we put absolute values everywhere that $\parslope$ and $\partrack$ appears, and multiply $1+w_{\rm de}$ by the sign, ${\rm sgn\,}(\parslope)={\rm sgn\,}(\partrack)$.

\subsection{Asymptotic Properties of $V(\phi)$ \& $\sqrt{\varepsilon_{\de \infty}}$ }

The class of quintessence/phantom models where the field is slow-rolling in the late-universe accelerating phase is of primary interest. Qualitatively this implies $\epsilon_{\de}\ll 1$ at low redshift,  but we would like to have a parametrization covering  a larger prior space allowing for higher $\epsilon_{\de}$ and $\vert d\psi /d\ln a \vert$,  and let observations determine the allowed speed of the roll. Therefore we also include moderate-roll models in our parametrization, making sure our approximate formula covers well trajectories with values of $\vert 1+w_{\de}\vert $ which extend up to $0.5$ and $\epsilon_{\de}$ to 0.75 at low redshift.

At high redshift the properties of the potential are poorly constrained by observations, and we have a very large set of possibilities to contend with. Since the useful data for constraining DE is at the lower redshifts, what we really need is just a reasonable shutoff at high redshift. One way we tried in early versions of this work was just to cap $\epsilon_{\de}$ at some  $\epsilon_{\de \infty}$ at a redshift well beyond the probe regime. Here we still utilize a cap as a parameter, but let physics be the guide to how it is implemented so that the $\epsilon_{\de}$ trajectories smoothly join higher to lower redshifts. 

The way we choose to do this here is to restrict our attention to tracking and thawing models for which the asymptotic forms are easily parameterized. Tracking models have an early universe attractor which implies  $\epsilon_{\de}$  indeed becomes a constant $\epsilon_{\de  \infty}$,  which is smaller than $\epsilon_m$.  If we apply this attractor  to equation~\refeq[odex], we  obtain $\sqrt{\epsilon_{V}\Omega_{\de}}$ is constant at high redshift, equaling our  $\sqrt{\varepsilon_{\de \infty}}$.   By definition, thawing models have $\epsilon_{\de  \infty}=0$. 

Consider the two high redshift possibilities exist for the tracking models. One has $\epsilon_{\de}= \epsilon_m$, which, when combined with equation~\refeq[odel], implies the potential structure parameter $\gamma$ vanishes, hence one gets  an an exponential potential as an asymptotic solution for $V$, as discussed in \citet{Copeland1998,Liddle1999} and \citet{Copeland2006}. 
Another possibility has $\epsilon_{\de}< \epsilon_m$, hence from Eq.~\refeq[odel] we obtain $\gamma = (\epsilon_m - \epsilon_{\de})/(2\epsilon_{\de})$ is a positive constant. Solving this equation for $\gamma$ yields a negative power-law potential, $V\propto \psi^{-1/\gamma}$. For both cases, we must have an asymptotically constant $\gamma\ge 0$ to give a constant \be
\varepsilon_{\de \infty} \rightarrow \epsilon_{m}/(1+2\gamma_{\infty} )  \ , \label{interp2}
\ee
where $\gamma_{\infty}$ is the high redshift limit of the shape parameter Eq.~\refeq[gammadef].
This also shows why $\varepsilon_{\de \infty} / \epsilon_{m}$ is actually a better parameter choice than $\varepsilon_{\de \infty}$, since the ratio is conserved as the matter EOS changes. 

The difference between tracking and thawing models is not only quantitative, but is also qualitative. For tracking models, the asymptotic limit $\sqrt{\varepsilon_{\de \infty}}$ has a dual interpretation. One is that the shape of potential has to be properly chosen to have the asymptotic limit of the right hand side of 
\be
  \sqrt{\varepsilon_{\de \infty}} \rightarrow \sqrt{\epsilon_V\Omega_{\de}}\vert_\infty = -\frac{1}{2} \frac{d\ln V}{d\psi}\vert_{\infty} \sqrt{\omphi_{\infty}}, \label{interp1}
\ee
existing. We already know how to choose the potential -- it has to be asymptotically either an exponential or a negative power-law. Another interpretation directly relates $\sqrt{\varepsilon_{\de \infty}} $ to the property of the potential through Eq.~\refeq[interp2]. In the thawing scenario, Eq.~\refeq[odel] is trivial, and the shape  parameter $\gamma$ is no longer tied to the vanishing  $\varepsilon_{\de \infty}$, though Eq.~\refeq[interp1] still holds. 

For phantom models the motivation for tracking solutions is questionable. We nevertheless allow for reciprocal $\epsilon_{\de}$-trajectories as for quintessence, just flipping the sign to extend the phenomenology to $1+w_{\de}<0$, as has become conventional in DE papers. What we do not do, however, is try to parameterize trajectories that cross  $\epsilon_{\de}=0$, as is done in the DETF $w_0$-$w_a$ phenomenology (see Fig.~\ref{figw0wa2D}). 

\subsection{The Two-Parameter $w_{\de}(a\vert \parslope , \partrack)$-Trajectories}

In the slow-roll limit, $\varepsilon_V$ does not vary much. As mentioned above, we use $\parslope \equiv \pm \epsilon_{V,eq} $ evaluated at the equality of matter and dark energy to characterize  
the (average) slope of $\ln V$ at low redshift, and $\Omega_{m0}$ or $h$ as a way to encode the actual value of the potential $V_0$ at zero redshift. To model $V(\phi)$ at high redshift for both tracking models and thawing models, we use the interpretation \refeq[interp1] characterized by  the  ``tracking parameter''  $\varepsilon_{\de \infty}= (\epsilon_V\Omega_{\de})\vert_{\infty}$, which is bounded by the tracking condition
$0 \le \varepsilon_{\de \infty}/\epsilon_m \le  1$. 

Eq.~\refeq[odex] shows that to solve for $\epsilon_{\de}(a)$ one only needs to know $\sqrt{\epsilon_V \Omega_{\de}}$ plus an initial condition,  $\sqrt{\varepsilon_{\de \infty}}$, but that too is just $\sqrt{\epsilon_V \Omega_{\de}}$ in the $a\rightarrow 0$ limit. We know as well  $\sqrt{\epsilon_V \Omega_{\de}}$ at $\aeq$, namely $\sqrt{\varepsilon_s /2}$. If the rolling is quite slow, $\epsilon_V$ will be nearly $\varepsilon_s$, and $\Omega_{\de}$ will be nearly $\Omega_{\rm \de app}=y^2$, hence 
\be
\sqrt{\epsilon_V \Omega_{\de}} \approx \sqrt{\varepsilon_s \Omega_{\rm \de app}}  = \sqrt{\varepsilon_s}y \ ,  
\ee
the 1-parameter approximation. But we are also assuming we know the $y=0 $ boundary condition, $\sqrt{\epsilon_V \Omega_{\de}}\vert_\infty$ as well as this $y_{\rm eq}$ value. If we make the simplest  linear-$y$ relation through the two points, we get our first order  approximation, Eq.~\refeq[lambdao1], for $\sqrt{\epsilon_V \Omega_{\de}} $. 

To get the DE EOS $w_{\de}$,  we need to integrate Eq.~\refeq[odex], with our $\sqrt{\epsilon_V \Omega_{\de}}$ approximation. In facilitate this, we make another approximation, 
\be
(\sqrt{\epsilon_V\Omega_{\de}} -\sqrt{\epsilon_{\de}} )  (1-\epsilon_{\de}/3) \approx (\sqrt{\epsilon_V\Omega_{\de}} -\sqrt{\epsilon_{\de}} ) \, , 
\ee
which is always a good one: at high redshift the tracking behaviour enforces $\sqrt{\epsilon_V\Omega_{\de}} -\sqrt{\epsilon_{\de}} \rightarrow 0$; and at low redshift $\epsilon_{\de}/3 \ll 1$. The analytic solution for $\sqrt{\epsilon_{\de}}$ retains the form linear in the two parameters, yielding Eq.~\refeq[eq:2paramsqeps], and hence the $w_{\de}$ approximation
\be
 1+w_{\de}(a)  \approx \frac{2}{3}[\sqrt{\varepsilon_{\de \infty}} + \left( \sqrt{\parslope} - \sqrt{2\varepsilon_{\de \infty}} \,  \right) F(\frac{a}{\aeq })]^2 ,  \label{twoparam}
\ee
where $F$ is defined in \refeq[Fxdef].

The three DE parameters $\aeq$, $\partrack$, and $\parslope$ are related to $\omm$ through the constraint equation
\be
\left[1+\exp{\left(2\int^1_{\aeq}(\epsilon_{\de}-\epsilon_m) \frac{da}{a}\right)}\right ]^{-1} = 1- \Omega_{m0}  \label{aeqconstraint}
\ee
obtained by integrating Eq.~\refeq[odeo] from $a=\aeq$, where by definition $\Omega_{\rm \de,eq}=1/2$, to today, $a=1$, hence $\Omega_{m0}$ is actually quite a complex parameter, involving the entire $w_{\de}(a;\aeq,\parslope,\partrack)/a$ history, as of course is $\aeq(\omm,\parslope,\partrack)$, which we treat as a derived parameter from the trajectories. 
The zeroth order solution for $\aeq(\omm,\parslope,\partrack)$ only depends upon $\Omega_{m0}$:  
\be
\aeq\approx \left(\frac{\Omega_{m0}}{1-\Omega_{m0}} \right)^{1/3} \ , \label{aeq0approx}
\ee
In conjunction with the approximate 2-parameter $w_{\de}$, \refeq[twoparam], this $\aeq$ completes our first approximation for DE dynamics. 

\subsection{The Three-parameter Formula}

The linear approximation of $\sqrt{\epsilon_V \Omega_{\de}}(y)$ and the zeroth order approximation for $\aeq$ rely on the slow-roll assumption. For moderate-roll models ($|1+w_{\de}|\gtrsim 0.2$), the two-parameter approximation is often not sufficiently accurate, with errors sometimes larger than $0.01$. We now turn to the improved 3-parameter fit  to $w_{\de}$; we need to considerably refine  $\aeq$ as well to obtain the desired high accuracy. 

The quadratic expansion Eq.~\refeq[3paramepsV] of$ \sqrt{\epsilon_V \Omega_{\de}}$ in $y$ leads to Eq.~\refeq[3parameps] for $\sqrt{\epsilon_{\de}}$, in terms of the two functions $F$ and $F_2$ of $a/\aeq$. Since $\parrun$ term is a ``correction term'', we impose a measure restriction on its uniform prior by requiring   $|\parrun| \lesssim 1$. Thus $w_\phi(a;\aeq,\parslope,\partrack,\parrun)$ follows, with the phantom paths covered by $\epsilon_{\rm \de,phantom}(a; \parslope , \partrack, \parrun )= {\rm sgn\,}(\parslope) \epsilon_{\rm \de,quintessence}(a; \vert\parslope \vert , \vert \partrack \vert, \parrun )$.   

We also need to improve $\aeq$, using the constraint Eq.~\refeq[aeqconstraint]. We do not actually need the exact solution, but just a good approximation that works for $\omm \sim 0.3$. For example, the following fitting formula is sufficiently good (error $\lesssim 0.01$) for $0.1<\omm<0.5$:
\bea
&&\ln \aeq(\omm,\parslope,\partrack,\parrun) = \frac{\ln \left[ \omm /(1-\omm)\right]}{3-{\rm sgn\,} (\parslope ) \delta} \ ,\label{aeqapp}
\eea
where  the correction to the index is 
\bea
\delta && \equiv   \left\{\sqrt{|\partrack|}+ \left[0.91-0.78\omm+ (0.24-0.76\omm)\parrun\right] \right. \nonumber \\
&& \times \left. \left(\sqrt{|\parslope|}-\sqrt{2|\partrack|}\right)\right\}^2   \nonumber \\
&& + \left[\sqrt{|\partrack|}+\left(0.53-0.09\parrun\right)\left(\sqrt{|\parslope|}-\sqrt{2|\partrack|}\right)\right]^2 . \label{deltadef} 
\eea

\section{Exact DE Paths for Various Potentials Compared with our Approximate Paths}

Equations~\refeq[3parameps] and (\ref{aeqapp}-\ref{deltadef}) define a three-parameter ansatz for $w_{\de} = -1+2/3\epsilon_\phi$ (with an implicit  fourth parameter for the energy scale, $\omm$). We numerically solve $w_\phi(a)$ for a wide variety of quintessence and phantom models, and show our $w_{\de}(a)$ formula follows the exact trajectories very well. This means we can compress this large class of theories into these few parameters.  

The dark energy parameters $\parslope$, $\partrack$ and $\parrun$ are calculated using definitions~\refeq[parslopedef], \refeq[partrackdef] and \refeq[parrundef]. We choose $z=50$ to calculate the high-redshift quantities such as $\varepsilon_V|_{a\ll 1}$. Unless otherwise specified, the initial condition is always chosen to be at $\ln a=-20$,  roughly at the time of BBN, at which point the initial field momentum is set to be zero (although it quickly relaxes). The figures express $\phi$ in units of the reduced Planck Mass $\mpl$, hence are $\sqrt{2}\psi$. The initial field value is denoted as $\phiini$.

\begin{figure}
\centering
\includegraphics[width=\narrowfigurewidth]{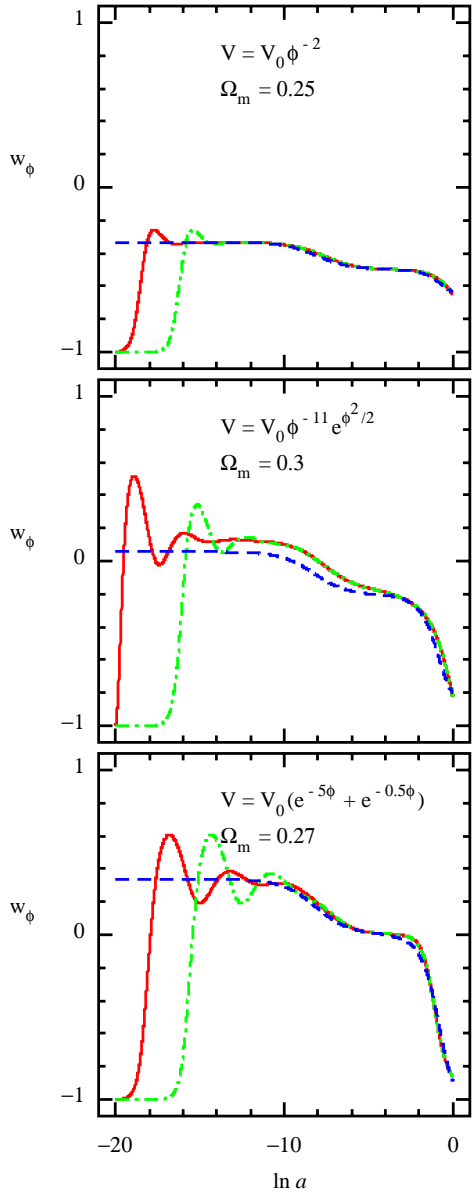} 
\caption{Examples of tracking models. The solid red lines and dot-dashed green lines are numerical solutions of $w_\phi$ with different $\phiini$. Upper panel: $\phi_{\rm ini; red} = 10^{-7}\mpl$, and $\phi_{\rm ini; green}=10^{-6}\mpl$. Middle panel: $\phi_{\rm ini; red}=0.01\mpl$, and $\phi_{\rm ini; green}=0.03\mpl$. Lower panel: $\phi_{\rm ini; red}=-12.5\mpl$, and $\phi_{\rm ini; green}=-10.5\mpl$. The dashed blue lines are $w(a)$ trajectories calculated with the three-parameter $w(a)$ ansatz \refeq[3parameps]. The rapid rise at the beginning is the $a^{-6}$ race of $\epsilon_\phi$ from our start of zero initial momentum towards the attractor.} \label{wfit_tracking}
\end{figure}

In the upper panel of Figure~\ref{wfit_tracking} we show that once the memory of  the initial input field momentum is lost, our parameterization fits the numerical solution for a negative power-law potential quite well over a vast number of e-foldings, even if $1+w$ is not small at low-redshift. 

We know that in order to achieve both $w_\phi \sim w_m$ at high redshift, which can alleviate the dark energy coincidence problem, and also slow-roll at low redshift, the negative power-law potential (or exponential potential) needs modification to fit the low redshift. One of the best known examples of a potential that does this is motivated by supergravity \cite{Brax1999}: 
\be
V(\phi)=V_0\left(\frac{\mpl}{\phi}\right)^{\alpha}\exp{\left(\frac{\phi^2}{2\mpl^2}\right)} \ , \label{sugra_potential}
\ee
where $\alpha\ge 11$. An example for a SUGRA model is shown in the middle panel of Figure~\ref{wfit_tracking}: it fits quite well for the redshifts over which we have data, with some deviation once $\epsilon_\phi$ exceeds unity at high redshift.  

Another popular tracking model is the double exponential model, an example of which is given in the lower panel of Figure~\ref{wfit_tracking}.

\begin{figure}
\centering
\includegraphics[width=\narrowfigurewidth]{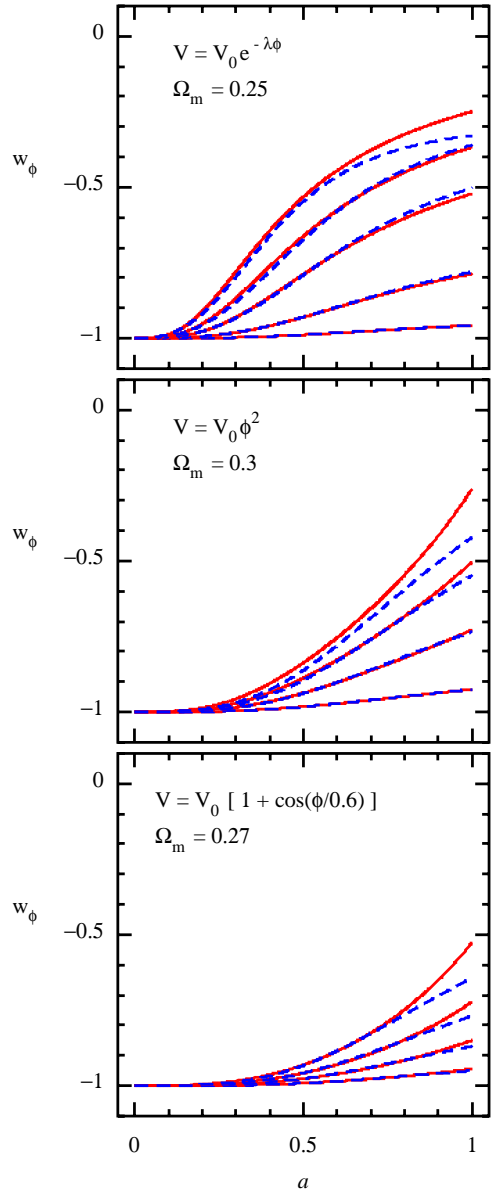} 
\caption{Examples of thawing models. The solid red lines are numerical solutions of $w_\phi$. The dashed blue lines are calculated using the three-parameter $w(a)$ formula \refeq[3parameps]. See the text for more details.} \label{wfit_thawing}
\end{figure}

In Figures~\ref{wfit_thawing} we show how robust our parametrization is for slow-roll and moderate-roll cases, taking examples from a variety of examples of popular thawing models. The horizontal axis is now chosen to be linear in $a$, since there is no interesting early universe dynamics in thawing models. The different realizations of the $w_{\de}$ trajectories are produced by choosing different values for the  potential parameters -- $\lambda$ for the upper panel, and $V_0$ for the middle and lower panels. The initial values of $\phi$ are chosen to ensure $\omphi (a=1)=1-\omm$ is satisfied. We see that in general our $w{\de}(a)$ parametrization works well up to $\vert 1+w \vert \sim 0.5$. The case shown in the bottom panel is the potential of a pseudo Nambu Goldstone boson, which has been much discussed for early universe inflation due to $\phi$ being an angular variable with a $2\pi$ shift symmetry that is easier to protect from acquiring large mass terms, and has been invoked for late universe inflation as well (e.g.,  \citet{Kaloper2006}). 

\begin{figure}
\centering
\includegraphics[width=\narrowfigurewidth]{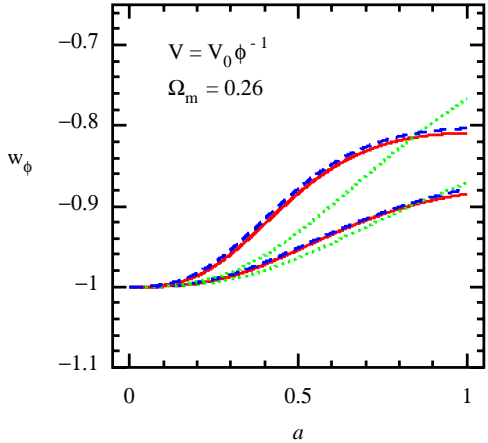}
\caption{An example showing how the $\parrun$ parameter improves our parametrization in the moderate-roll case. The solid red lines are numerical solutions of $w_\phi$. The dashed blue lines are $w(a)$ trajectories calculated with the three-parameter ansatz \refeq[3parameps]. The dotted green lines are two-parameter approximations obtained by forcing $\parrun =0$.} \label{wfit_thawing_2and3}
\end{figure}

In Figure~\ref{wfit_thawing_2and3} we demonstrate how the parameter $\parrun$ improves our parametrization to sub-percent level. In the slow-roll regime, the $\parrun$ correction is small, and both the two- and three-parameter formulas can fit the numerical solution well. But the $\parrun$ correction becomes important in the moderate-roll regime.

\begin{figure}
\centering
\includegraphics[width=\narrowfigurewidth]{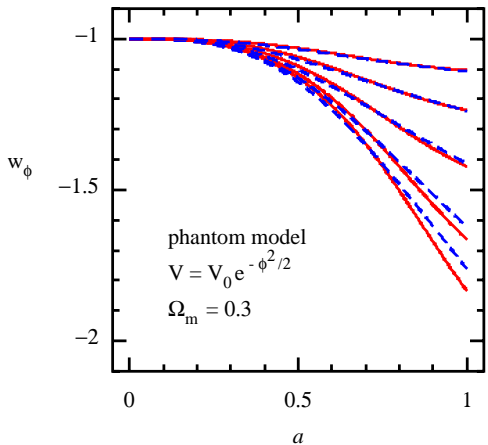}
\caption{An example of phantom model. The solid red lines are numerical solutions of $w_\phi$. The dashed blue lines are $w(a)$ trajectories calculated with the three-parameter $w(a)$ formula \refeq[3parameps].} \label{wfit_phantom}
\end{figure}

The last example, shown in Figure~\ref{wfit_phantom}, is a phantom model. 

\section{Observational Constraints}\label{sec:constraint}
In this section we compile the updated cosmological data sets and use them to constrain the quintessence and phantom models.

\subsection {Current Data Sets Used}

For each of the data sets used in this paper we either wrote a new module to calculate the likelihood or modified the CosmoMC likelihood code to include dynamic $w$ models.

\begin{enumerate}

\item[]{\it Cosmic Microwave Background (CMB)}

Our complete CMB data sets include WMAP-7yr \cite{Komatsu2010,Jarosik2010}, ACT \cite{Fowler2010}, BICEP \cite{Chiang2009}, QUaD \cite{Castro2009}, ACBAR \cite{Reichardt2008,Kuo2006,Runyan2003, Goldstein2003}, CBI \cite{Sievers2007,Readhead2004a,Readhead2004b,Pearson2003}, BOOMERANG \cite{Jones2006,Piacentini2006,Montroy2006}, VSA \cite{Dickinson2004}, and MAXIMA \cite{Hanany2000}. 

For high resolution CMB experiments we need to take account of other sources of power beyond the primary CMB.  We always include the CMB lensing contribution even though with current data its influence is not yet strongly detected \cite{Reichardt2008}. For most high resolution data sets, radio sources have been subtracted and residual contributions have been marginalized over \cite{Fowler2010}. However, other frequency-dependent sources will be lurking in the data, and they too should be marginalized over. We follow a well-worn path for treating the thermal Sunyaev-Zeldovich (SZ) secondary anisotropy \cite{Sunyaev1972,Sunyaev1980}: we use a power template from a gasdynamical cosmological simulation (with heating only due to shocks) described in  \cite{Bond2002,Bond2005} with an overall amplitude multiplier $A_{\rm SZ}$, as was used in the CBI and ACBAR papers. This template does not differ by much from the more sophisticated ones obtained by  \citet{Battaglia2010} that include cooling and feedback. Other  SZ template choices with different shapes and amplitudes  \cite{Komatsu2002, Sehgal2009}  have been made by WMAP and ACT and SPT, and as a robustness alternative by CBI. A key result is that as long as the ``nuisance parameter'' $A_{\rm SZ}$ is marginalized, other cosmological parameters do not vary much as the SZ template changes. We make no other use of $A_{\rm SZ}$ here, although it can encode the tension between  the cosmology derived from the CMB primary anisotropy and the predictions for the SZ signal in that cosmology. 

In addition to thermal SZ, there are a number of other sources: kinetic SZ with a power spectrum whose shape roughly looks like the thermal SZ one \cite{Battaglia2010}; sub-mm dusty galaxy sources, which have clustering contributions in addition to Poisson fluctuations.  Thermal SZ is a small effect at WMAP and Boomerang resolution. At CBI's 30 GHz, kinetic SZ is small compared with thermal SZ, but it is competitive at ACBAR, QUaD's and ACT's 150 GHz and of course dominates at the $\sim$220 GHz  thermal-SZ null, but the data is such that little would be added by separately modelling the frequency dependence and shape difference of it, so de facto it is bundled into the generic $A_{\rm SZ}$ of the frequency-scaled thermal-SZ template. We chose not to include the SPT data in our treatment because the influence of the  sub-mm dusty galaxy sources should be simultaneously modelled, and where its band-powers lie (beyond $\ell \sim 2000$), the primary SZ power is small.

\item[]{\it Type Ia Supernova (SN)}

We use the \citet{Kessler2009} data set, which combines the SDSS-II SN samples with the data from the ESSENCE project \cite{Miknaitis2007, Wood-Vasey2007}, the Supernova Legacy Survey (SNLS) \cite{Astier2006}, the Hubble Space Telescope (HST) \cite{Garnavich1998, Knop2003,Riess2004,Riess2007}, and a compilation of nearby SN measurements. Two light curve fitting methods, MLCS2K2 and SALT-II, are employed in \citet{Kessler2009}. The different assumptions about the nature of the SN color variation lead to a significant apparent discrepancy in $w$. For definiteness,  we choose the SALT-II-fit results in this paper, but caution that if we had tried to assign a systematic error to account for the differences in the two methods the error bars we obtain would open up, and the mean values for $1+w$ may not centre as much around zero as we find. 

\item[]{\it Large Scale Structure (LSS)}

The LSS data are for the power spectrum of the Sloan Digital Sky Survey Data Release 7 (SDSS-DR7) Luminous Red Galaxy (LRG) samples \cite{Reid2009}. We have modified the original LSS likelihood module to make it compatible with time-varying $w$ models.

\item[]{\it Weak Lensing (WL)}

 Five WL data sets are used in this paper. The effective survey area $A_{\textrm{eff}}$ and galaxy number density $n_{\textrm{eff}}$ of each survey are listed in Table~\ref{tblwldata}.

\begin{deluxetable}{lll}
\tabletypesize{\scriptsize}
\tablewidth{0pt}
\tablecaption{Weak Lensing Data Sets}
\tablehead{ \colhead{ Data sets} & \colhead{$A_{\textrm{eff}}$ (deg$^2$)}  & \colhead{ $n_{\rm eff}$  (arcmin$^{-2}$)} }
\startdata
  COSMOS\tablenotemark{a}  & 1.6 & 40\\
  CFHTLS-wide\tablenotemark{b} & 22 & 12 \\
  GaBODS\tablenotemark{c} & 13 & 12.5 \\
  RCS\tablenotemark{d} & 53 & 8 \\
  VIRMOS-DESCART\tablenotemark{e} & 8.5 &  15 
\enddata
  \tablenotetext{a}{\citet{Massey2007,Lesgourgues2007}.}
  \tablenotetext{b}{\citet{Hoekstra2006,Schimd2007}.}
  \tablenotetext{c}{\citet{Hoekstra2002a,Hoekstra2002b}.}
  \tablenotetext{d}{\citet{Hoekstra2002a,Hoekstra2002b}.}
  \tablenotetext{e}{\citet{Van-Waerbeke2005,Schimd2007}.}
\label{tblwldata}
\end{deluxetable}

For the COSMOS data, we use the CosmoMC plug-in written by Julien Lesgourgues \cite{Lesgourgues2007} with our modifications for dynamic dark energy models.

For the other four weak lensing data sets we use the covariance matrices given by \citet{Benjamin2007}. To calculate the likelihood we wrote a CosmoMC plug-in code. We take the best fit parameters $\alpha,\beta,z_0$ for $n(z)\propto (z/z_0)^\alpha \exp{\left[-(z/z_0)^\beta\right]}$, and marginalize over $z_0$, assuming a Gaussian prior with a width such that the mean redshift $z_m$ has an uncertainty of $0.03(1+z_m)$. We have checked that further marginalizing over other $n(z)$ parameters ($\alpha$ and $\beta$) has no significant impact \citep{Amigo2008}.

\item[]{\it Lyman-$\alpha$ Forest (Ly$\alpha$)}

Two Ly$\alpha$ data sets are applied: i) the set from \citet{Viel2004} consisting of the LUQAS sample \cite{Kim2004} and the data given in \citet{Croft2002}; ii) the SDSS Ly$\alpha$ data presented in \citet{McDonald2005} and \citet{McDonald2006}. To calculate the likelihood we interpolate the $\chi^2$ table in a three dimensional parameter space, where the three parameters are amplitude, index, and the running of linear CDM power spectrum at pivot $k=0.9 h$ Mpc$^{-1}$.

\item[]{\it Other Constraints}

We have also used in CosmoMC the following observational constraints: i) the distance-ladder constraint on Hubble parameter, obtained by the Hubble Space Telescope (HST) Key Project \cite{Riess2009}; ii) constraints from the Big Bang Nucleosynthesis (BBN): $\Omega_{b0} h^2 = 0.022\pm 0.002 $ (Gaussian prior) and $\Omega_\Lambda (z=10^{10})<0.2$ \citep[see][]{Steigman2006,O'Meara2006,Ferreira1997,Bean2001,Copeland2006}; iii) an isotopic constraint on the age of universe $10\text{Gyr}<\text{Age}<20\text{Gyr}$ \citep[see e.g.][]{Dauphas2005}. For the combined data sets, none of these add much. They are useful if we are looking at the impact of the various data sets in isolation on constraining our parameters. 

\end{enumerate}

\subsection{CosmoMC Results for Current Data}\label{sec:cosmomc}

Using our modified CosmoMC, we ran Markov Chain Monte Carlo (MCMC) calculations to determine the likelihood of cosmological parameters, which include the standard six, $A_{\rm SZ}$ and our DE parameters, $w_0$-$w_a$ for Fig.~\ref{figw0wa2D} and the new ones $\parslope$, $|\partrack|/\epsilon_m$ and $\parrun$ for most of the rest. Here we use $|\partrack|/\epsilon_m$ as a fundamental parameter to eliminate the dependence of $\partrack$ on $\epsilon_m$ and the sign of $\parslope$, whereas $\partrack$ adjusts as one evolves from a relativistic to non-relativistic matter EOS. The main results are summarized in Table~\ref{tblcosmomc}. The basic six are: $\Omega_{b0}h^2$, proportional to the current physical density of baryons; $\Omega_{c0}h^2$, the physical cold dark matter density; $\theta$, the angle subtended by sound horizon at ``last scattering'' of the CMB, at $z\sim 1100$; $\ln A_s$, with $A_s$ the primordial scalar metric perturbation power evaluated at pivot wavenumber $k=0.002 {\rm Mpc}^{-1}$; $n_s$, the spectral index of primordial scalar metric perturbation; $\tau$,  the reionization Compton depth. The measures on each of these variables is taken to be uniform. Derived parameters include $z_{\rm re}$,  the reionization redshift; Age/Gyr, the age of universe in units of gigayears; $\sigma_8$, today's amplitude of the linear-extrapolated matter density perturbations over an  top-hat spherical window with radius $8 h^{-1}{\rm Mpc}$. And, of great importance for DE, $\omm$;  and $H_0$ in unit of \hunit, which set the overall energy scale of late-inflatons.  

\begin{figure}
\centering
\includegraphics[width=\narrowfigurewidth]{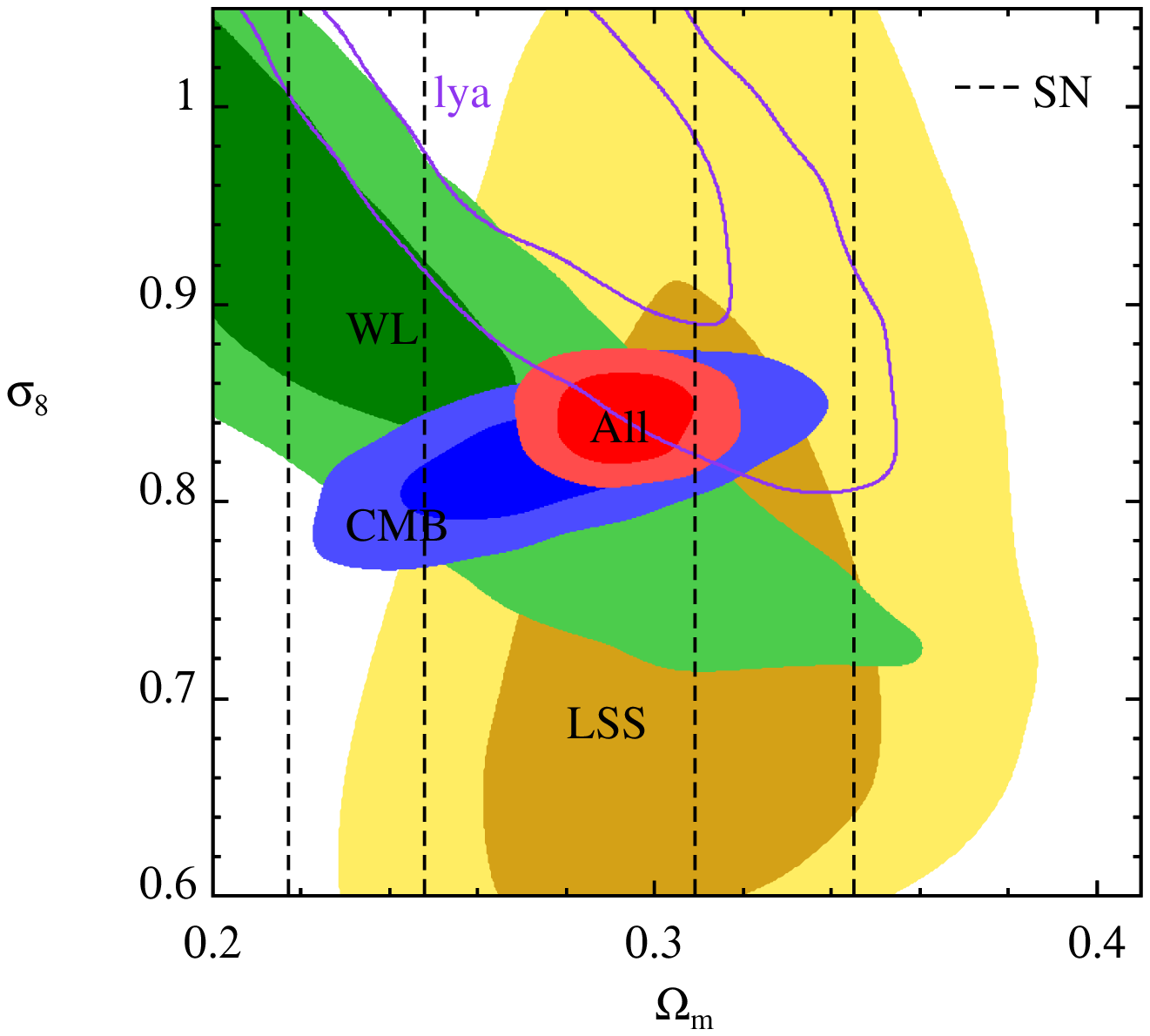}
\caption{The marginalized 68.3\% CL and 95.4\% CL constraints on $\sigma_8$ and $\omm$ for the $\Lambda$CDM model vary with different choices of data sets. For each data set a HST constraint on $H_0$ and BBN constraint on $\Omega_{b0}h^2$ have been used.}\label{sigma8_omm}
\end{figure}

\begin{figure}
\centering
\includegraphics[width=\narrowfigurewidth]{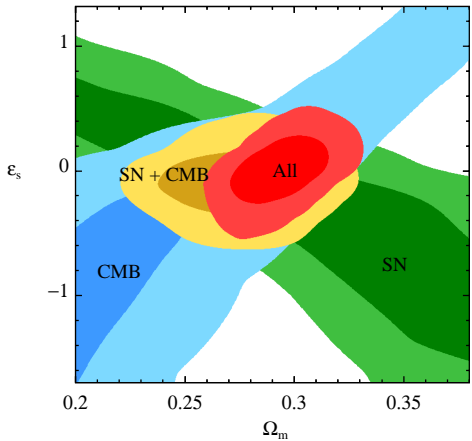}
\caption{The marginalized 68.3\% CL and 95.4\% CL constraints on $\parslope$ and $\omm$, using different (combinations) of data sets. This is the key DE plot for late-inflaton models of energy scale, encoded by $1-\omm$ and potential gradient defining the roll-down rate, $\sqrt{\parslope}$. }\label{epsCombo2D}
\end{figure}

We begin with our versions of the familiar contour plots that show how the various data sets combine to limit the size of the allowed parameter regions.  Figures~\ref{sigma8_omm} and \ref{epsCombo2D} shows the best-fit cosmological parameters do depend on the specific subset of the data that is used.  We find most cosmological parameters are stable when we vary the choice of data sets, The data making the largest differences are:  the SDSS-DR7-LRG data driving $\Omega_m$ up, and the Ly$\alpha$ data driving  $\sigma_8$ up. We take SN+CMB as a ``basis'', for which $\omm =0.267^{+0.019}_{-0.018}$ and $\sigma_8=0.817^{+0.023}_{-0.021}$. Adding LSS pushes $\omm$ to $0.292^{+0.011}_{-0.010}$; whether or not WL and Ly$\alpha$ are added or not is not that relevant, as Figure~\ref{sigma8_omm} shows. Ly$\alpha$ pushes $\sigma_8$ to $0.844^{+0.015}_{-0.016}$. These drifts of best-fit values are at the  $\sim 2 \sigma$, level in terms of the $\sigma$'s derived from using all of the data sets, as is evident visually in Figure~\ref{sigma8_omm}.

\tabletypesize{\normalsize}
\setlength{\tabcolsep}{0.04in}
\begin{deluxetable*}{cccccc}
\tablecaption{Cosmic Parameter Constraints: $\Lambda$CDM, $w_0$-CDM, $w_0$-$w_a$-CDM, $\parslope$-$\partrack$-$\parrun$-CDM }
\tablehead{\colhead{} &  \colhead{$ \Lambda $CDM} & \colhead{$w=w_0$}  & \colhead{$w$=$w_0$+$w_a(1-a)$} & \colhead{track+thaw} & \colhead{thaw}} \\
\startdata
\multicolumn{6}{c}{Current Data: CMB+LSS+WL+SN1a+Ly$\alpha$} \\ 
\hline
$ \Omega_{b0}h^2 $ & $0.0225^{+0.0004}_{-0.0004}$ & $0.0225^{+0.0004}_{-0.0004}$ & $0.0225^{+0.0004}_{-0.0004}$ & $0.0225^{+0.0004}_{-0.0004}$ & $0.0225^{+0.0004}_{-0.0004}$ \\
 $ \Omega_{c0}h^2 $ & $0.1173^{+0.0020}_{-0.0019}$ & $0.1172^{+0.0024}_{-0.0024}$ & $0.1177^{+0.0025}_{-0.0023}$ & $0.1174^{+0.0029}_{-0.0029}$ & $0.1175^{+0.0023}_{-0.0022}$ \\
 $ \theta $ & $1.042^{+0.002}_{-0.002}$ & $1.042^{+0.002}_{-0.002}$ & $1.042^{+0.003}_{-0.002}$ & $1.042^{+0.003}_{-0.002}$   & $1.0417^{+0.0021}_{-0.0021}$   \\
 $ \tau $ & $0.089^{+0.014}_{-0.013}$ & $0.090^{+0.015}_{-0.014}$ & $0.088^{+0.014}_{-0.013}$ & $0.090^{+0.015}_{-0.014}$& $0.088^{+0.015}_{-0.014}$  \\
 $ n_s $ & $0.96^{+0.01}_{-0.01}$ & $0.96^{+0.01}_{-0.01}$ & $0.96^{+0.01}_{-0.01}$ & $0.96^{+0.01}_{-0.01}$& $0.958^{+0.011}_{-0.011}$  \\
 $ \ln(10^{10}A_s) $ & $3.24^{+0.03}_{-0.03}$ & $3.24^{+0.03}_{-0.03}$ & $3.24^{+0.03}_{-0.03}$ & $3.24^{+0.03}_{-0.03}$  & $3.239^{+0.033}_{-0.033}$ \\
 $ A_{SZ} $ & $0.56^{+0.11}_{-0.15}$ & $0.56^{+0.11}_{-0.14}$ & $0.57^{+0.11}_{-0.14}$ & $0.56^{+0.11}_{-0.14}$& $0.56^{+0.11}_{-0.14}$ \\
 $ \Omega_m $ & $0.292^{+0.011}_{-0.010}$ & $0.294^{+0.012}_{-0.012}$ & $0.293^{+0.012}_{-0.012}$ & $0.293^{+0.015}_{-0.014}$ & $0.293^{+0.013}_{-0.011}$  \\
 $ \sigma_8 $ & $0.844^{+0.015}_{-0.016}$ & $0.841^{+0.026}_{-0.026}$ & $0.847^{+0.026}_{-0.026}$ & $0.844^{+0.035}_{-0.036}$ & $0.844^{+0.024}_{-0.023}$  \\
 $ z_{re} $ & $10.8^{+1.2}_{-1.1}$ & $10.9^{+1.2}_{-1.2}$ & $10.7^{+1.1}_{-1.1}$ & $10.9^{+1.2}_{-1.2}$  & $10.8^{+1.2}_{-1.2}$   \\
 $ H_0 $ & $69.2^{+1.0}_{-1.0}$ & $69.0^{+1.4}_{-1.4}$ & $69.2^{+1.4}_{-1.4}$ & $69.0^{+1.9}_{-1.8}$  & $69.1^{+1.4}_{-1.4}$   \\
  $ w_0 $ & \nodata & $-0.99^{+0.05}_{-0.06}$ & $-0.98^{+0.14}_{-0.11}$ & \nodata &\nodata  \\
 $ w_a $ & \nodata & \nodata & $-0.05^{+0.35}_{-0.58}$ & \nodata & \nodata \\
 $ \varepsilon_s $ & \nodata & \nodata & \nodata & $0.00^{+0.18}_{-0.17}$ & $-0.00^{+0.27}_{-0.29}$ \\
 $ |\partrack|/\epsilon_m $ & \nodata & \nodata & \nodata & $0.00^{+0.21+0.58}$ & \nodata  \\
 $ \zeta_s $ & \nodata & \nodata & \nodata & n.c. & n.c. \\
\hline
 \multicolumn{6}{c}{\,}\\
 \multicolumn{6}{c}{Forecasted Data: Planck2.5yr + low-$z$-BOSS + CHIME + Euclid-WL + JDEM-SN} \\
\hline
 $ \Omega_{b0}h^2 $ & $0.02200^{+0.00007}_{-0.00007}$ &
$0.02200^{+0.00007}_{-0.00007}$ & $0.02200^{+0.00007}_{-0.00008}$ &
$0.02200^{+0.00007}_{-0.00008}$ & $0.02200^{+0.0007}_{-0.0007}$\\
$ \Omega_{c0}h^2 $ & $0.11282^{+0.00024}_{-0.00023}$ &
$0.11280^{+0.00027}_{-0.00027}$ & $0.11282^{+0.00026}_{-0.00029}$ &
$0.1128^{+0.0003}_{-0.0003}$ & $0.1128^{+0.0003}_{-0.003}$\\
$ \theta $ & $1.0463^{+0.0002}_{-0.0002}$ &
$1.0463^{+0.0002}_{-0.0002}$ & $1.0463^{+0.0003}_{-0.0002}$  &
$1.0463^{+0.0002}_{-0.0002}$ & $1.0463^{+0.0002}_{-0.0002}$ \\
$ \tau $ & $0.090^{+0.003}_{-0.003}$ & $0.090^{+0.004}_{-0.004}$ &
$0.090^{+0.004}_{-0.004}$ & $0.090^{+0.005}_{-0.005}$ &
$0.090^{+0.004}_{-0.004}$ \\
$ n_s $ & $0.970^{+0.002}_{-0.002}$ & $0.970^{+0.002}_{-0.002}$ &
$0.970^{+0.002}_{-0.002}$ & $0.970^{+0.002}_{-0.002}$ &
$0.970^{+0.002}_{-0.002}$ \\
$ \ln(10^{10}A_s) $ & $3.115^{+0.008}_{-0.008}$ &
$3.115^{+0.009}_{-0.009}$ & $3.115^{+0.009}_{-0.009}$  &
$3.115^{+0.010}_{-0.010}$ & $3.115^{+0.009}_{-0.009}$\\
$ \Omega_m $ & $0.260^{+0.001}_{-0.001}$ & $0.261^{+0.002}_{-0.002}$
& $0.260^{+0.003}_{-0.003}$ & $0.2609^{+0.0022}_{-0.0022}$ &
$0.2605^{+0.0027}_{-0.0024}$  \\
 $ \sigma_8 $ & $0.7999^{+0.0016}_{-0.0017}$ &
$0.7994^{+0.0023}_{-0.0025}$ & $0.800^{+0.003}_{-0.003}$  &
$0.7992^{+0.0027}_{-0.0027}$ & $0.7996^{+0.0026}_{-0.0029}$\\
$ z_{re} $ & $10.9^{+0.3}_{-0.3}$ & $10.9^{+0.3}_{-0.3}$ &
$10.9^{+0.3}_{-0.3}$   & $10.9^{+0.4}_{-0.4}$ & $10.9^{+0.3}_{-0.3}$ \\
$ H_0 $ & $72.0^{+0.1}_{-0.1}$ & $71.9^{+0.3}_{-0.3}$ &
$72.0^{+0.4}_{-0.4}$ & $71.85^{+0.37}_{-0.29}$
&$71.94^{+0.34}_{-0.36}$ \\
 $ w_0 $ & \nodata & $-1.00^{+0.01}_{-0.01}$ &
$-1.00^{+0.03}_{-0.03}$    & \nodata &\nodata \\
 $ w_a $ & \nodata & \nodata & $0.01^{+0.08}_{-0.08}$  &\nodata & \nodata \\
$ \varepsilon_s $ & \nodata & \nodata & \nodata &
$0.005^{+0.031}_{-0.025}$ & $0.008^{+0.056}_{-0.054}$\\
 $ |\partrack|/\epsilon_m $ & \nodata & \nodata & \nodata & $0.000^{+0.034+0.093}$
&\nodata \\
$ \zeta_s $ & \nodata & \nodata & \nodata  & n.c. & n.c. 
\enddata
\tablenotetext{}{Tracking + thawing models use $w_{\rm de}(a\vert \parslope, \partrack, \parrun )$ of Eq.~\refeq[3parameps]. Thawing enforces $\partrack=0$. n.c. stands for ``not constrained''. $H_0$ has units \hunit. $\theta$ is in $\mathrm{radians}/100$. $\epsilon_m$ is 3/2 in the matter-dominated regime. $\partrack = 3(1+w_{\de \infty})/2$. }
\label{tblcosmomc}
\end{deluxetable*}

\begin{deluxetable}{ll}
\tablewidth{0pt}
\tabletypesize{\scriptsize}
\tablecaption{The marginalized 68.3\%, 95.4\%, and 99.7\% CL constraints on $\parslope$ under different prior assumptions.}
\tablehead{\colhead{Prior} & \colhead{Constraint}}
\startdata
flat prior on $\parslope$ & $\parslope = 0.00^{+0.18+0.39+0.72}_{-0.17-0.44-0.82}$ \\
flat prior on $\sqrt{|\parslope|}$ & $\parslope = 0.00^{+0.09+0.27+0.46}_{-0.07-0.28-0.52}$ \\
thawing prior $\partrack=0$ & $\parslope = 0.00+^{0.27+0.53+0.80}_{-0.29-0.61-0.98}$ \\
slow-roll thawing $\partrack=\parrun=0$  & $\parslope=-0.01^{+0.26+0.50+0.70}_{-0.28-0.59-0.86}$ \\
quintessence $\parslope>0$ &  $\parslope = 0.00 ^{+0.18+0.39+0.76}$ 
\enddata
\label{tblpriors}
\end{deluxetable}

\begin{figure}
\centering
\includegraphics[width=\halffigurewidth]{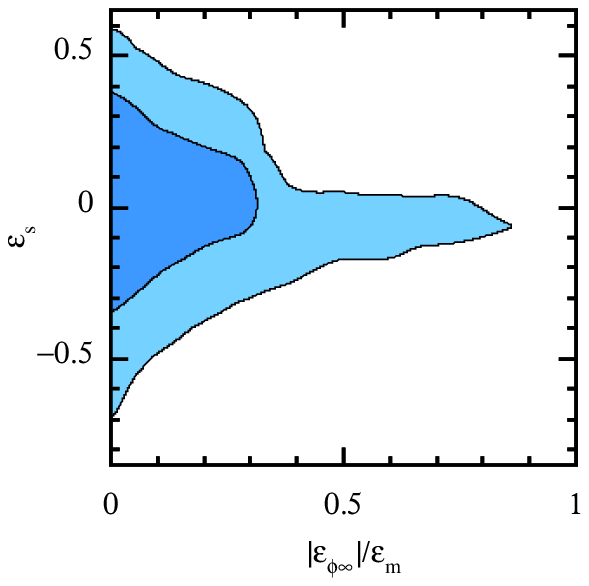}%
\includegraphics[width=\halffigurewidth]{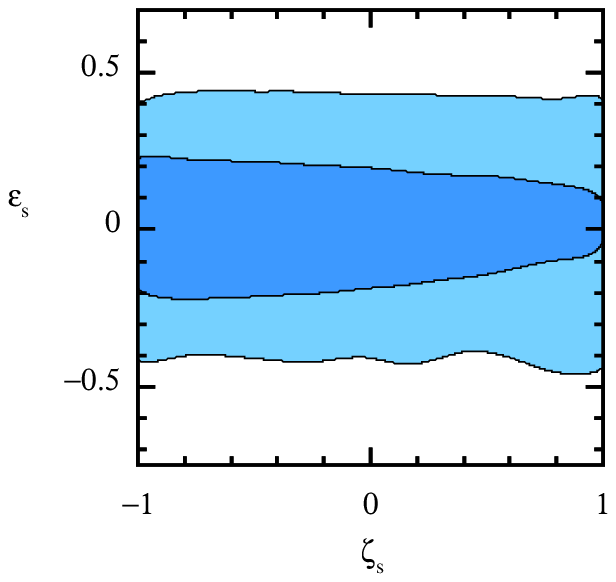}
\caption{Marginalized 2D likelihood contours for our three DE parameters derived using ``ALL'' current observational data. The inner and outer contours are 68.3\%  CL and 95.4\% CL contours. (We actually show $|\partrack| /\epsilon_m$ since that is the attractor whether one is in the relativistic $\epsilon_m=2$ or non-relativistic $\epsilon_m =3/2$ regime.) }\label{contours2D}
\end{figure}

Figure~\ref{epsCombo2D} shows how the combination of complementary data sets constrains $\parslope$ and $\omm$, the two key parameters that determine low-redshift observables. The label ``All'' refers to all the data sets described in this section, and ``CMB'' refers to all CMB data sets, and so forth. In Figure~\ref{contours2D}, marginalized 2D likelihood contours are shown for all our DE parameters. The left panel shows that the slope parameter $\parslope$ and tracking parameter $\partrack$ are both constrained. The constraint on $\parslope$ limits how steep the potential could be at low redshift. The upper bound of $\partrack$ indicates that the field can not be rolling too fast at intermediate redshift $z\sim 1$. The right panel shows that $\parrun$ is not constrained, and is almost uncorrelated with $\parslope$. This is because the current observational data favors slow-roll ($|1+w_{\de}|\ll 1$), in which case the $\parrun$ correction in $w_\phi$ is very small. Another way to interpret this is that a slowly rolling field does not ``feel'' the curvature of potential. In Section~\ref{sec:discussion} we will discuss the meaning and measurement of $\parrun$ in more detail.

Because of the correlation between $\parslope$ and $\partrack$, the marginalized likelihood of $\parslope$ depends on the prior of $\partrack$. On the other hand, the constraint on $\parslope$ will also depend on the prior on $\parslope$ itself. For example, we can apply a flat prior on $d\ln V/d\phi|_{a=\aeq}$, rather than a flat prior on the squared slope $\varepsilon_s$. Other priors we have tried are the ``thawing prior'' $\partrack = 0$, the ``quintessence prior'' $\parslope >0$, and ``slow-roll thawing prior'' $\partrack = \parrun=0$. The results are summarized in Table~\ref{tblpriors}. 

\begin{figure*}
\centering
\includegraphics[width=\halffigurewidth]{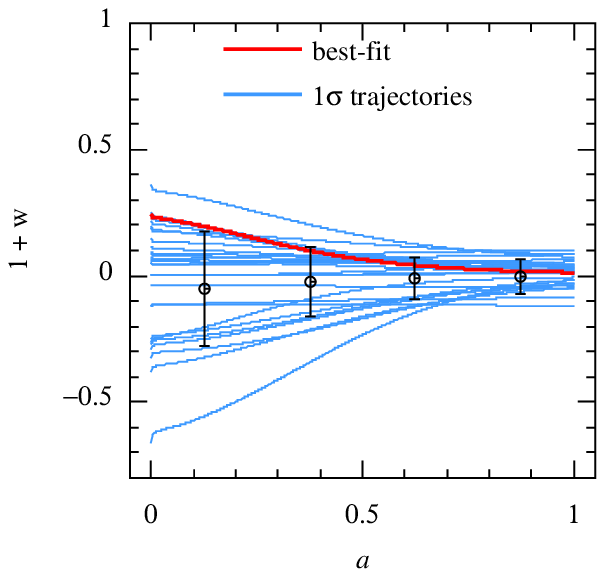}%
\includegraphics[width=\halffigurewidth]{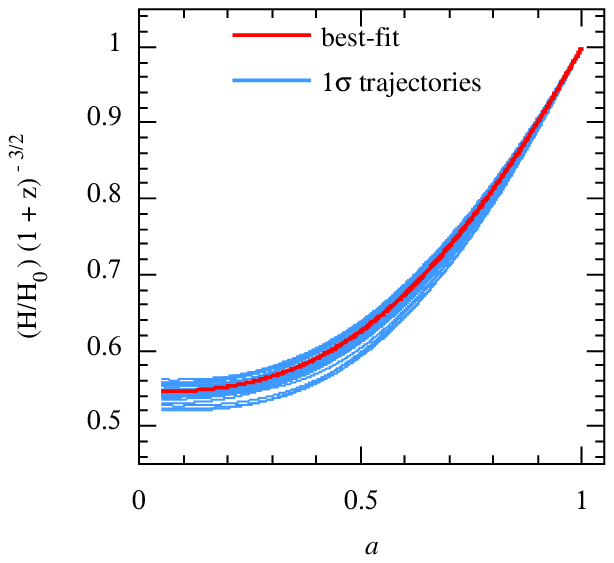}%
\includegraphics[width=\halffigurewidth]{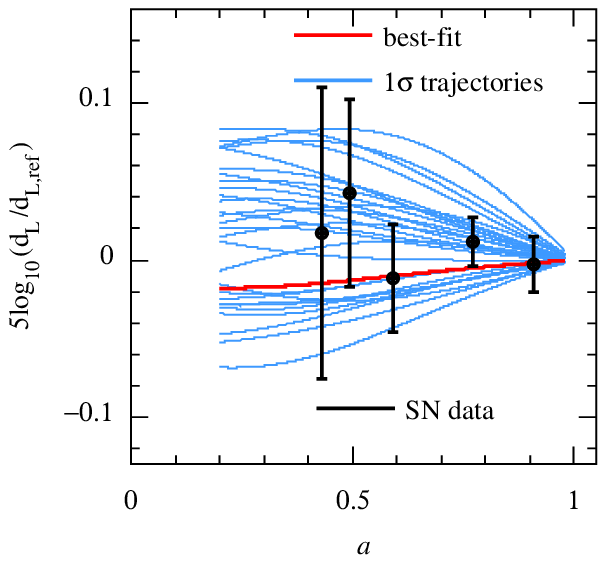}%
\includegraphics[width=\halffigurewidth]{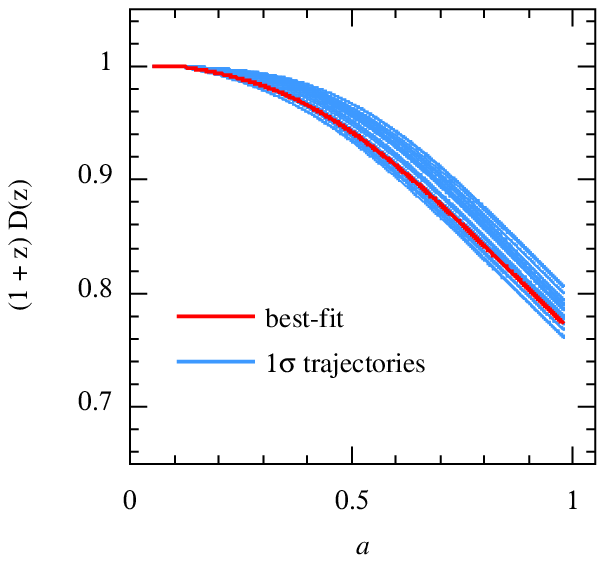}
\includegraphics[width=\halffigurewidth]{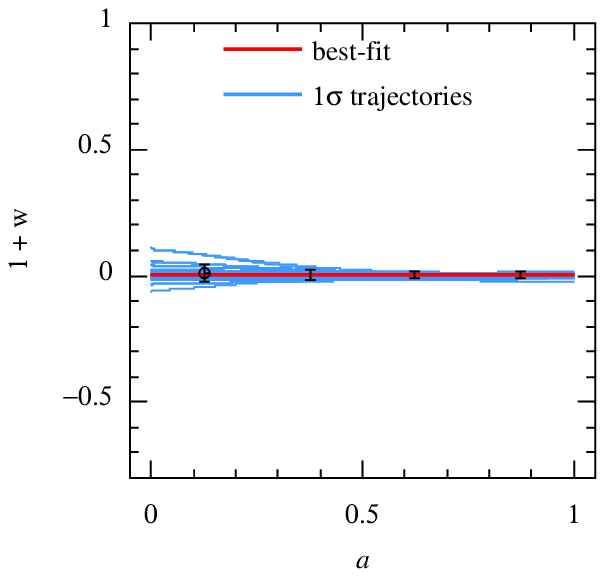}%
\includegraphics[width=\halffigurewidth]{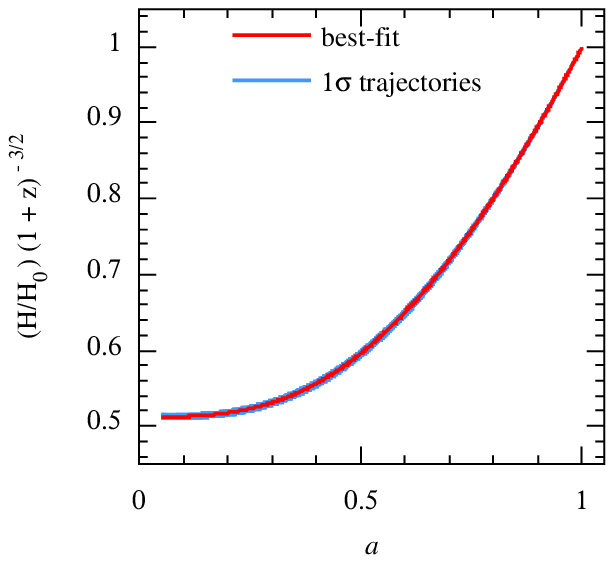}%
\includegraphics[width=\halffigurewidth]{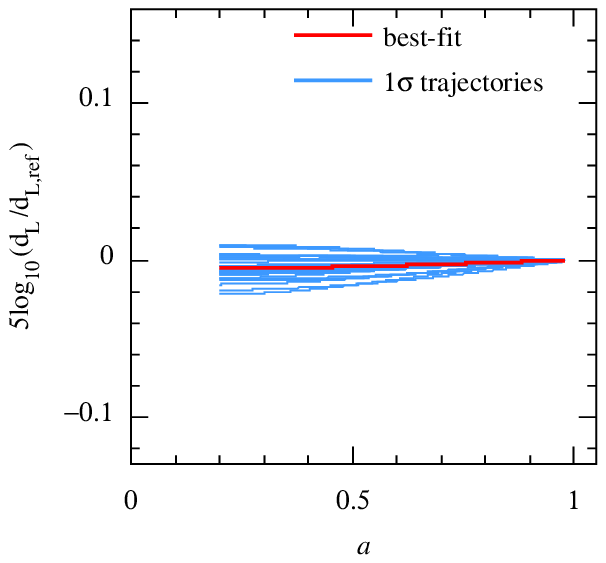}%
\includegraphics[width=\halffigurewidth]{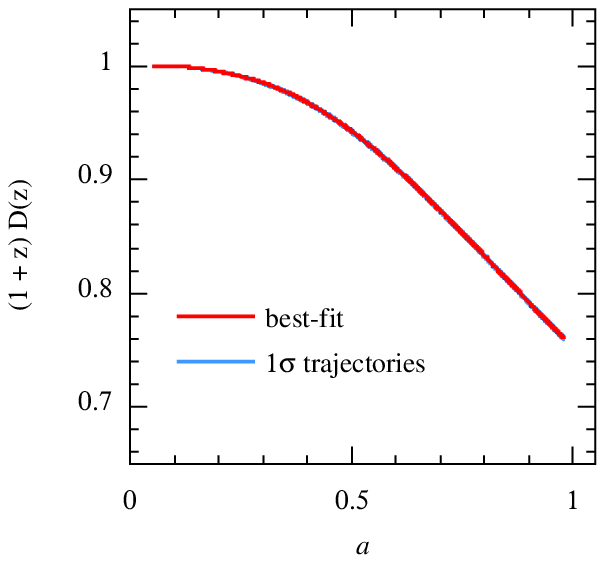}
\caption{The best-fit trajectory (heavy curve) and a sample of trajectories that are within one-sigma (68.3\% CL). In the upper panels the current data sets are used, and lower panels the forecast mock data. From left to right the trajectories are the dark energy equation of state, the Hubble parameter rescaled with $H_0^{-1}(1+z)^{3/2}$, the distance moduli with a reference $\Lambda$CDM model subtracted, and the growth factor of linear perturbation rescaled with a factor $(1+z)$ (normalized to be unit in the matter dominated regime). The current supernova data is plotted agianst the reconstructed trajectories of distance moduli. The error bars shown in the upper-left and lower-left panels are one-$\sigma$ uncertainties of $1+w$ in bands $0<a\le 0.25$, $0.25<a\le0.5$, $0.5<a\le0.75$ and $0.75<a\le1$. } \label{figtrajs}
\end{figure*}

In Figure~\ref{figtrajs} we show the reconstructed trajectories of the dark energy EOS, Hubble parameter, distance moduli, and the growth factor $D$ of linear perturbations. The $w_{\de}$ information has also been compressed into a few bands with errors, although that is only to guide the eye. The off-diagonal correlation matrix elements between bands is large, encoding the coherent nature of the trajectories. As well, the  likelihood surface in all $w_{\de}$-bands  is decidedly non-Gaussian, and should be characterized for this information to be statistically useful in constraining models by itself. The other variables shown involve integrals of $w_{\de}$, and thus are not as sensitive to detailed rapid change aspects of $w_{\de}$, which, in any case, our late-inflaton models do not give. The bottom panels show how impressive the constraints on trajectory bundles should become with planned experiments, a subject to which we now turn. 

\section{Future Data Forecasts}\label{sec:forecast}

In this section ,we discuss the prospects for further constraining the parameters of the new $w(a)$ parameterization using a series of forthcoming or proposed cosmological observations: from the Planck satellite CMB mission \cite{Tauber2005,Clavel2005, book_PlanckBlue}; from a JDEM  (Joint Dark Energy Mission, \citet{Albrecht2006}) for Type Ia supernova observations; from a weak lensing survey by the Euclid satellite \cite{book_EUCLIDYellow};  and from  future BAO data that could be obtained by combining low-redshift galaxy surveys with a  redshifted-21-cm survey of moderate $z\sim 2$. The low-redshift galaxy surveys for BAO information can be achieved by combining a series of ground-based galaxy observations, such as BOSS \cite{Schlegel2007}. For the 21-cm survey, we assume a $200{\rm m}\times 200{\rm m}$ ground-based cylinder radio telescope \cite{Chang2007, Peterson2009, Morales2009}, which is the prototype of the proposed experiment CHIME (Canadian Hydrogen Intensity Mapping Experiment).

\begin{deluxetable}{llllll}
\tabletypesize{\scriptsize}
\tablewidth{0pt}
\tablecaption{Fiducial model used in future data forecasts}
\tablehead{\colhead{$\Omega_{b0}h^2$} & \colhead{$\Omega_{c0}h^2$} & \colhead{$h$} & \colhead{ $\sigma_8 $} & \colhead{ $n_s$} &\colhead{ $\tau$ }}
\startdata
$0.022$ & $0.1128$  & $0.72$  & $0.8$  & $0.97$  & $0.09$ 
\enddata
\label{tblfid}
\end{deluxetable}

\subsection{The mock data sets}

\subsubsection{Planck CMB simulation}

Planck 2.5 years (5 sky surveys) of multiple (CMB) channel data are used in the forecast, with the instrument characteristics for the channels used listed in Table~\ref{tblPlanck} using Planck ``Blue Book'' detector sensitivities and the values given for the full width half maxima.

\begin{deluxetable}{llll}
\tabletypesize{\scriptsize}
\tablewidth{0pt}
\tablecaption{Plank  Instrument Characteristics }
\tablehead{\colhead{Channel Frequency (GHz)} & \colhead{70} &\colhead{100} & \colhead{143}}
\startdata
Resolution\tablenotemark{a} (arcmin) & 14 & 10 & 7.1 \\
Sensitivity \tablenotemark{b}  - intensity ($\mu K$)&  8.8 & 4.7 & 4.1 \\
Sensitivity - polarization ($\mu K$)& 12.5 & 7.5  & 7.8 
\enddata
\tablenotetext{a}{Full width at half maximum (FWHM) assuming Gaussian beams.} 
\tablenotetext{b}{This is for 30 months of integration.}
\vspace{0.03in}
\label{tblPlanck}
\end{deluxetable}

For a nearly full-sky (we use $f_{\rm sky}=0.75$) CMB experiment, the likelihood ${\cal L}$ can be approximated with the following formula \citep{Baumann2008}:
\bea
-2 \ln{{\cal L}} &=& \sum_{l=l_{\rm min}}^{l_{\rm max}} (2l+1)f_{\rm sky} \nonumber \\
&& \left[ -3 + \frac{\hat{C}_l^{BB}}{C_l^{BB}} +  \ln\left(\frac{C_l^{BB}}{\hat{C}_l^{BB}}\right)  \right. \nonumber \\
&& + \frac{\hat{C}_l^{TT}C_l^{EE} + \hat{C}_l^{EE}C_l^{TT} - 2\hat{C}_l^{TE}C_l^{TE}}{C_l^{TT}C_l^{EE}-(C_l^{TE})^2}  \nonumber \\
&&\left. +\ln{\left(\frac{C_l^{TT}C_l^{EE}-(C_l^{TE})^2}{\hat{C}_l^{TT}\hat{C}_l^{EE}-(\hat{C}_l^{TE})^2}\right)}\right] \ , 
\eea
where $l_{\rm min}=3$ and $l_{\rm max}=2500$ have been used in our calculation. Here, $\hat{C}_l$ is the observed (or simulated input) angular power spectrum, and $C_l$ is the theoretical power spectrum plus noise.

We use the model described in \citet{Dunkley2008} and \citet{Baumann2008} to propagate the effect of polarization foreground residuals into the estimated uncertainties on the cosmological parameters. For simplicity, only the dominating components in the frequency bands we are using, i.e., the synchrotron and dust signals, are considered in our simulation. The fraction of the residual power spectra are all assumed to be 5\%. 

\subsubsection{JDEM SN simulation}

For the JDEM SN simulation, we use the model given by the Dark Energy Task Force (DETF) forecast \cite{Albrecht2006}, with roughly 2500 spectroscopic supernova at $0.03<z<1.7$ and 500 nearby samples. The apparent magnitude of SN is modelled as
\begin{eqnarray}
m &=& M+5\log_{10}\left(\frac{d_L}{\text{Mpc}}\right)+25 \nonumber \\
&& -\mu^Lz-\mu^Qz^2 - \mu^S\delta_{\text{near}}\ , \label{snm}
\end{eqnarray}
where $\delta_{\text{near}}$ is unity for the nearby samples and zero otherwise.

There are four nuisance parameters in this model. The supernova absolute magnitude is expanded as a quadratic function $M-\mu^Lz-\mu^Qz^2$ to account for the possible redshift dependence of the SN peak luminosity, where $M$ is a free parameter with a flat prior over $-\infty<M<+\infty$; and for $\mu^L$ and $\mu^Q$, Gaussian priors $\mu^{L,Q}=0.00\pm 0.03/\sqrt{2}$ (the ``pessimistic'' case in the DETF forecast) are applied. Finally, given that the nearby samples are obtained from different projects, an offset $\mu^S$ is added to the nearby samples only. For $\mu^S$ we apply a Gaussian prior $\mu^S=0.00\pm 0.01$.

The intrinsic uncertainty in the supernova absolute magnitude is assumed to be $0.1$, to which an uncertainty due to a peculiar velocity $400$~km/s is quadratically added.

\subsubsection{BAO simulation}

The ``Baryon Acoustic Oscillations'' (BAO) information can be obtained by combining a series of ground-based low-redshift galaxy surveys with a high-redshift 21-cm survey. 
We assume a fiducial galaxy survey with comoving galaxy number density $0.003h^3{\rm Mpc}^{-3}$ and sky coverage 20,000 deg$^2$, which is slightly beyond, but not qualitatively different from the specification of SDSS-III BOSS (Baryon Oscillation Spectroscopic Survey) project. The 21-cm BAO survey using a ground-based cylinder radio telescope has been studied by \citet{Chang2007, Peterson2009} and \citet{Seo2009}. The specifications we have used are listed in Table~\ref{tbl21cmspec}. 

\begin{deluxetable}{ll}
\tablewidth{0pt}
\tabletypesize{\scriptsize}
\tablecaption{21-cm BAO Survey Specifications}
\tablehead{\colhead{Parameter} & \colhead{Specification} }
\startdata
shot noise & $0.01 h^3 {\rm Mpc}^{-3}$ \\
survey area & 15,000 deg$^2$ \\
number of receivers & 4000 \\
integration time & 4 years \\
cylinder telescope & $200$ m $\times 200$ m \\
antenna temperature & $50 K$ \\
bias &  1 
\enddata
\label{tbl21cmspec}
\end{deluxetable}

For more details about the BAO forecast technique, the reader is referred to \citet{Seo2007} and \citet{Seo2009}.

\subsubsection{EUCLID Weak Lensing Simulation} \label{subsubsec:dune_wl}

We assume a weak lensing survey with the following ``Yellow Book'' EUCLID specifications \cite{book_EUCLIDYellow}:
\be 
f_{\text{sky}}=0.5 \, ,\  \langle \gamma_{\text{int}}^2 \rangle^{1/2}=0.35 \, ,\bar{n}=40 \text{ galaxies/arcmin}^{2} \, , 
\ee
where $\langle \gamma_{\text{int}}^2 \rangle^{1/2}$ is the intrinsic galaxy ellipticity, and $\bar{n}$ is the average 
galaxy number density being observed. At high $\ell$, the non-Gaussianity of the dark matter density field becomes important \cite{Dore2009}. For simplicity we use a $\ell$-cutoff $\ell_{\rm max}=2500$ to avoid modelling the high-$\ell$ non-Gaussianity, which will provide more cosmic information, hence the weak lensing constraints shown in this paper are conservative.

We use a fiducial galaxy distribution,  
\begin{equation}
  n(z)\propto(\frac{z}{z_0})^2\exp{\left[-\left(\frac{z}{z_0}\right)^{1.5}\right]}, \ {\rm with} \ z_0=0.6. 
\end{equation}
The galaxies are divided into four tomography bins, with the same number of galaxies in each redshift bin. 
The uncertainty in the  median redshift in each redshift bin is assumed to be $0.004(1+z_m)$, where $z_m$ is the median redshift in that bin. 

In the ideal case in which the galaxy redshift distribution function is perfectly known, the formula for calculating weak lensing tomography 
observables and covariance matrices can be found in e.g. \citet{Ma2006}. In order to propagate the uncertainties of photo-z parameters onto the uncertainties of the cosmological parameters, 
we ran Monte Carlo simulations to obtain the covariance matrices due to redshift uncertainties, which are then added to the ideal covariance matrices. 

\subsection{Results of the Forecasts}

The constraints on cosmological parameters from future experiments are  shown in Table~\ref{tblcosmomc} below those for current data. The future observations should improve the measurement of  $\parslope$ significantly -- about five times better than the current best constraint. There is also a significant improvement on the upper bound of $\partrack$, which together with the constraint on $\parslope$ can be used to rule out many tracking models. The running parameter $\parrun$ remains unconstrained.

\begin{figure}
\centering
\includegraphics[width=\narrowfigurewidth]{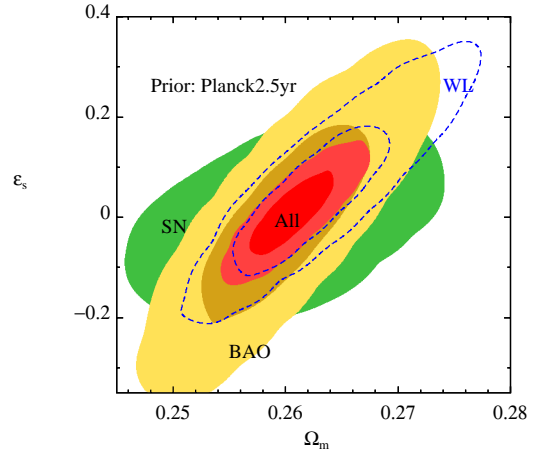} 
\caption{Marginalized 2D likelihood contours; Using mock data and assuming thawing prior ($\partrack=0$); The inner and outer contours of each color correspond to 68.3\% CL and 95.4\% CL, respectively. See the text for more details.}\label{figcomboforecasts}
\vspace{0.05in}
\end{figure}

To compare different dark energy probes, in Figure~\ref{figcomboforecasts} we plot the marginalized constraints on $\Omega_{m0}$ and $\parslope$ for different data sets. To break the degeneracy between dark energy parameters and other cosmological parameters, we apply the Planck-only  CMB constraints as a prior on each of the non-CMB data sets.

The forecasted BAO, WL, and SN results, when combined with the Planck prior, are all comparable. We note in particular that the ground-based BAO surveys deliver similar measurements at a fraction of the cost of the space experiments, although of course much useful collatoral information will come from all of the probes. 

The shrinkage in the allowed parameter space is visualized in Figure~\ref{figtrajs} through the decrease in area of the trajectory bundles from current to future data, showing  trajectories of $1+w_{\de}$ sampled down to the one-sigma level.  We also show the mean and standard deviation of $1+w$ at the center of four uniform bands in $a$. As mentioned above, the bands are highly correlated because of the coherence of the trajectories, a consequence of their physical origin. We apply such band analyses to early (as well as late) universe inflation that encodes such smoothing theory priors in future papers.

Figure~\ref{figtrajs} also plots $(1+z)^{-3/2}(H/H_0) = \sqrt{\rho_{\rm tot}a^3/\rho_{\rm tot,0}}$ trajectories, which are flat in the high redshift matter-dominated regime. They are less spread out than the $w$ bundle because $H$ depends upon an integral of $w$. One of the observable quantities measured with supernovae is the luminosity distance. What we plot is something more akin to relative magnitudes of standard candles, $5\log_{10}\left(d_L/d_{\rm L; ref}\right)$, in terms of  $d_{\rm L;ref}$,  the luminosity distance of a reference $\Lambda$CDM model. For luminosity distance trajectories reconstructed with the current data we choose a reference model with $\omm=0.29$, which the SDSS LSS data drove us to;  for forecasts, we chose $\omm=0.26$, what the other current data is more compatible with,  for the reference. Because the supernova data only measure the ratio of luminosity distances, for each trajectory we normalize $d_L/d_{\rm L;ref}$ to be unit in the low-redshift limit by varying $H_0$ in the reference model. The error bars shown are for the current supernova data \cite{Kessler2009}, showing compatibility with $\Lambda$CDM, and, based upon the coherence of the quintessence-based prior and the large error bars, little flexibility in trying to fit the rise and fall of the data means. 

The rightmost panels of Figure~\ref{figtrajs} shows the linear growth factor for dark matter fluctuations relative to the expansion factor, $(1+z)D(z)$. The normalization is such that it is unity in the matter-dominated regime. It should be determined quite precisely for the quintessence prior with future data. 

\section{Discussion and Conclusions} \label{sec:discussion}

\subsection{Slow-roll Thawing Models, Their One-parameter Approximation and the Burn-in to It}

\begin{figure}
\plottwo{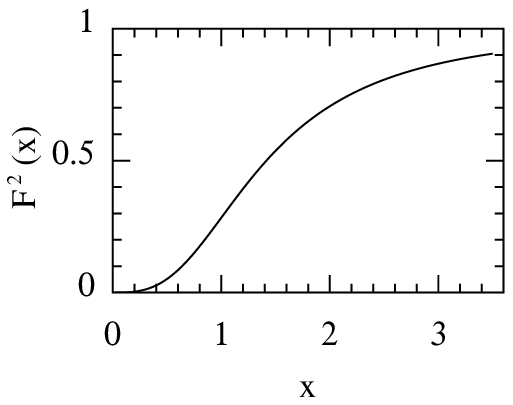}{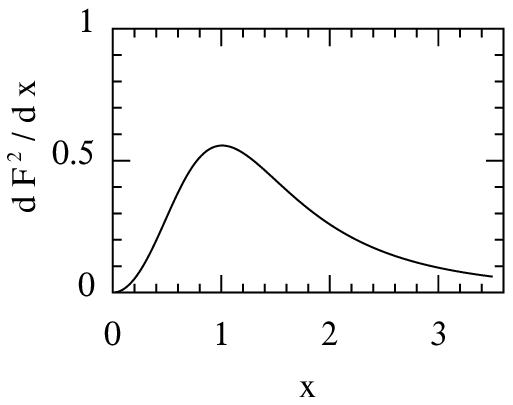} 
\caption{Left panel: the function $F^2(x)$ with $x=a/\aeq$ defined by Eq.~\refeq[Fxdef]. Right Panel: the derivative $dF^2/dx$, showing where $w_{\phi}$ changes most quickly in thawing models, namely near $\aeq$.} \label{figFofx}
\end{figure}

In slow-roll thawing models, to  first order  $w(z)$ only depends on two physical quantities: one determining ``when to roll down'', quantified mostly by $1-\Omega_{m0}$, and one determining ``how fast to roll down'', quantified by the slope of the potential, {\it i.e.}, $\varepsilon_s$, since $\partrack=0$, and $\parrun \approx 0$: 
\be
1+w_\phi \vert_{\rm slowroll, thawing} \approx \frac{2\varepsilon_s}{3} F^2(\frac{a}{\aeq})\, , \label{oneparam}
\ee
 where $F$ is an analytical function given by Eq.~\refeq[Fxdef], and $\aeq \vert_{\rm slowroll, thawing} \approx \left(\frac{\omm}{1-\omm}\right) ^{1/\left[3-1.08\left(1-\omm\right)\parslope\right]}$ to sufficient accuracy. We plot $F^2(x)$ in the left panel of Figure~\ref{figFofx}. It is zero in the small $a$ regime and unity in the large $a$ regime, and at $\aeq$ is $F^2(1)=\left[\sqrt{2}-\ln(1+\sqrt{2})\right]^2 \approx 0.284$.  The right panel of Figure~\ref{figFofx} plots the derivative $dF^2/dx$ showing $|dw/da|$ maximizes around  $a=\aeq$ (at redshift $\sim 0.4$).  Three stages are evident in the significant time-dependence of $dw/da$: the Hubble-frozen stage at $a\ll \aeq$, where $w$ has the asymptotic value $-1$; the thawing stage when the dark energy density is comparable to the matter component; and the future inflationary stage where dark energy dominates ($a\gg \aeq$) and a future attractor with $1+w\rightarrow  2\parslope /3$ is approached, agreeing with the well-known early-inflation solution $1+w= 2\epsilon_V/3$.

Even with such models, there is in principle another parameter: the initial field momentum is unlikely to be exactly on the $\partrack =0$ attractor. In that case, $\dot{\phi}$ falls like $a^{-3}$ until the attractor is reached, a ``burn-in'' phase. (Such burn-in phases are evident in Fig.~\ref{wfit_tracking} for tracking models.)  In early work on parameterizing late-inflatons we added a parameter $a_s$ to characterize when this drop towards the attractor occurred, and found it could be (marginally) constrained by the data to be small relative to the $a$-region over which the trajectories feel the DE-probing data. However, we decided to drop this extra parameter for this paper since the expectation of late-inflaton models is that the attractor would have been established long before the redshift range relevant for DE-probes. 

\begin{figure}
\centering
\includegraphics[width=\narrowfigurewidth]{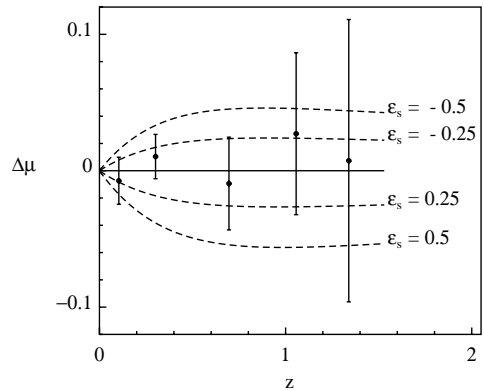} 
\caption{The dependence of distance moduli $\mu(z)$ on the slope parameter $\parslope$ for slow-roll thawing models. The prediction of $\mu(z)$ from a reference $\Lambda$CDM model with $\omm=0.29$ is substracted from each line. The supernova samples are binned into redshift bins with bin width $\Delta z = 0.3$. 
}\label{eps_SN}
\end{figure}

The 3-parameter, 2-parameter and 1-parameter fits all give about the same result for the statistics of $\parslope$ derived from current data.   
Figure~\ref{eps_SN} illustrates why we find $\parslope \approx 0.0\pm 0.28$. It shows the confrontation of banded supernova data on distance moduli defined by 
\be
\mu\equiv 25+ 5\log_{10}{\left( \frac{d_L}{\rm Mpc}\right )} \ , \label{mudef}
\ee
where $d_L$ is the physical luminosity distance, with trajectories in $\mu$ with differing values of $\parslope$ (for the $h$ (or $\omm$) values as described.) The current DE constraints are largely determined by  SN, in conjunction with the $\Omega_{m}h^2$-fixing CMB. 

\subsection{A $w_0$-$w_a$ Degeneracy for the Slow-roll Thawing Model Prior} \label{sec:w0wadegeneracy}
 
The parameterization~\refeq[oneparam] can be simply related to the phenomenological $w_0$-$w_a$ parameterization by a first order Taylor series expansion about a redshift zero pivot: 
\be
1+w_0\equiv \frac{2}{3} \varepsilon_s F^2(\frac{1}{\aeq}),
\ee
and
\be  w_a\equiv - \frac{2}{3} (\varepsilon_s /\aeq )\left. \frac{dF^2(x)}{dx}\right\vert_{x=1/\aeq} .
\ee
From formula~\refeq[oneparam],  it follows that  $w_0$ and $w_a$ should satisfy the linear relation
\be
1+w_0+w_a(1-a^*)=0 \ , \label{slowroll_constr1}
\ee
where
\be
a^*=1-\frac{\sqrt{1+\aeq ^3}-\aeq ^3 \ln \left(\frac{1+\sqrt{1+\aeq ^3}}{\aeq ^{3/2}} \right)}{6\aeq ^3 \left[ \ln\left(\frac{1+\sqrt{1+\aeq
^3}}{\aeq ^{3/2}}\right)-\frac{1}{\sqrt{1+\aeq^3}} \right]}
\ee
is roughly the scale factor where the field is unfrozen. Over  the range of $0.17<\Omega_{m}<0.5$, this  constraint equation~\refeq[slowroll_constr1] can be well-approximated by the formula
\be
1+w_0+w_a \left(0.264+\frac{0.132}{\omm} \right)=0\ . \label{slowroll_constr}
\ee
Such a degeneracy line is plotted in Figure~\ref{figw0wa2D}. Rather than a single $w_0$ parameter, constraining to this line based on a more realistic theoretical prior obviously makes for a more precise measurement of $w$. On the other hand, once the measurements of $w_0$ and $w_a$ are both sufficiently accurate, the slow-roll thawing models could be falsified if they do not encompass part of the line. Obviously though it would be better to test deviations relative to true thawing trajectories rather than the ones fit by this perturbation expansion about redshift zero. One way to do this is to demonstrate that $1+w_\infty$ is not compatible with zero, as we now discuss. 

\subsection{Transforming $w_0$-$w_a$ and $\parslope$-$\partrack$ Contours into $w_0$-$w_\infty$ Contours } \label{sec:w0winfty}

\begin{figure}
\centering
\includegraphics[width=\figurewidth]{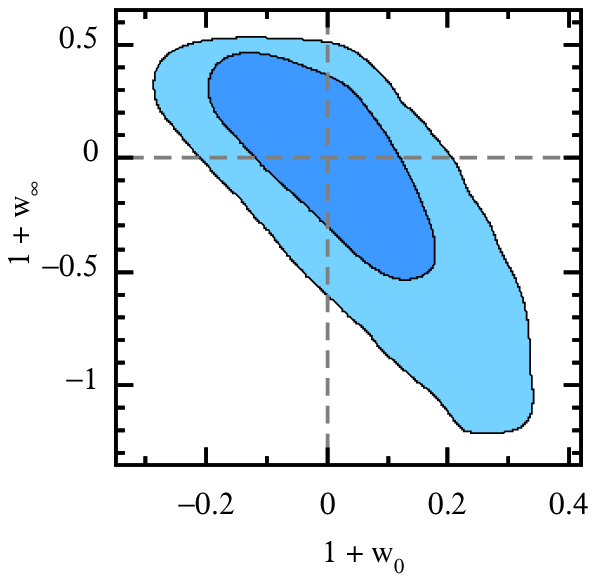}
\includegraphics[width=\figurewidth]{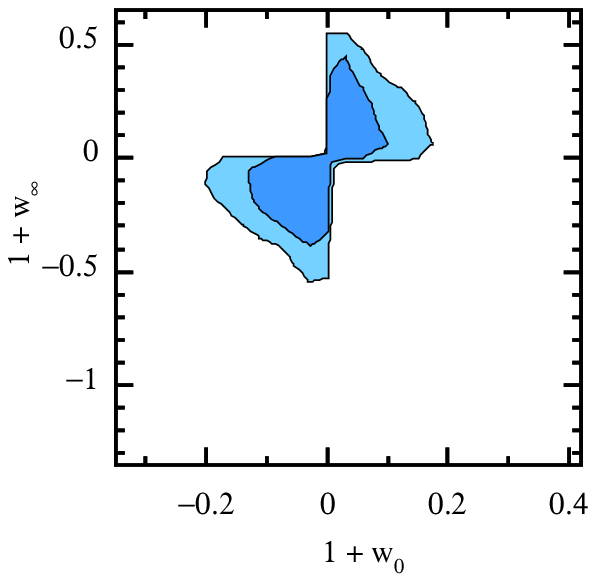}
\caption{The upper panel shows the marginalized 68.3\% (inner contour) and 95.4\% (outer contour) constraints on $w_0$ and $w_\infty$ for the conventional linear DETF parametrization, recast as $w=w_\infty +(w_0 - w_\infty ) a$, using the current data sets described in \S~\ref{sec:constraint}. It is a slightly tilted version of the $w_0$-$w_a$ version in Fig.~\ref{figw0wa2D}. The demarcation lines transform to just the two axes, with the upper right quadrant the pure quintessence regime, and the lower right quadrant the pure phantom regime. If the cross-over of partly phantom and partly quintessence are excluded, as they are for our late-inflaton treatment in this paper, the result looks similar to the lower panel, in which the 2-parameter $\parslope$-$\partrack$ formula has been recast into $2\varepsilon_0 /3$ and $2\varepsilon_\infty /3$, via $\sqrt{\varepsilon_0} =  \sqrt{\partrack}  +(\sqrt{\parslope}- \sqrt{2\partrack})F(1/\aeq )$.  However, as well the prior measure has also been transformed, from our standard uniform $d\parslope d\partrack$  one to a uniform  $d\varepsilon_0 d\partrack$ one. The quadrant exclusions are automatically included in our re-parameterization.}\label{figw0winfty2D}
\end{figure}

The DETF $w_0$-$w_a$ parameterization which is linear in $a$ can be thought of as a parameterization in terms of  $w_0$ and $w_\infty = w_a + w_0$ (or $\epsilon_{\rm de,0}$ and $\epsilon_{\rm de, \infty}$ in the DE acceleration factor language).  The upper panel $w_0$-$w_a$ contour map in  Fig.~\ref{figw0winfty2D} shows that it is only a minor adjustment of the $w_0$-$w_a$ of Fig.~\ref{figw0wa2D}. If we said there is a strict dichotomy between trajectories that are quintessence and trajectories that are phantom, as has been the case in our treatment in this paper, only the two of the four quadrants of Fig.~\ref{figw0winfty2D} would be allowed. (In the $w_0$-$w_a$ figures, the prior looks quite curious visually, in that the line $w_a=-(1+w_0)$ through the origin has only the region above it allowed in the $1+w_0 >0$ regime, and only the region below it allowed in the $1+w_0 <0$ regime.)

To make this issue more concrete, we show in the lower  panel of Fig.~\ref{figw0winfty2D} what happens for our physics-motivated 2-parameter linear model, Eq.~\ref{twoparam}, except it is linear for $\pm \sqrt{\vert 1+w_{\rm de} \vert}$, and in the ``time variable'' $F$ not in $a$. In terms of  $w_{\de 0}$ and $w_{\de \infty}$, we have $\sqrt{\epsilon_{\de }} =  \sqrt{\epsilon_{\de 0}} + ( \sqrt{\epsilon_{\de \infty}}  - \sqrt{\epsilon_{\de 0}})\left[1-F(a/\aeq )/F(1/\aeq )\right]$. The prior measure on $1+w_{\de 0}$ and $w_{\de \infty}$ are uniform as in the linear-$a$ case. Once the prior that only 2 quadrants are allowed is imposed upon the linear-$a$ case, it looks reasonably similar to the lower panel. 

To illustrate that the determination of our various parameters actually depends upon an extended redshift regime, one can construct window functions,  that is redshift-filters, for the parameters. These determine how an error in a parameter is built up by a sum of deviations of the observational data from the observable constructed using the model for $w_{\rm de}$, using the maximum likelihood formula. This shows that the window functions are quite extended, not concentrated around zero redshift for $w_0$,  not concentrated at high redshift for $w_\infty$ or $\partrack$ and not concentrated at $\aeq$ for $\parslope$.  An illustration of the redshift reach of the $\parslope$ parameter is
Fig.~\ref{eps_SN} for the Supernova observable, namely a magnitude difference. The specific window functions are very data-dependent of course. 

\subsection{$\parrun$, the Potential Curvature  and the Difficulty of  Reconstructing $V(\phi)$}

For slow-roll thawing quintessence models, the running parameter $\parrun$ can be related to the second derivative of $\ln V$. Using  the single-parameter approximation we can approximately calculate $d\phi/d\ln a$ at $a=\aeq$. The result is:
\be
\left.\frac{d\phi}{d\ln a}\right\vert_{a=\aeq}= \left. \sqrt{2} \mpl\sqrt{\epsilon_{\de} \Omega_{\de}}\right\vert_{a=\aeq} \approx 0.533 \sqrt{\parslope}\mpl \ . \label{dphidN_approx}
\ee 
An immediate consequence is that the amount that $\phi$ rolls, at least at late time, is small compared with the Planck mass. 
In the slow-roll limit, to the zeroth order, Eq.~\refeq[3paramepsV] can be reformulated as
\be
\frac{d\sqrt{2\varepsilon_V}}{dN}\vert_{a=\aeq}\approx \frac{3}{2\sqrt{2}}\parrun\sqrt{\parslope} \ . \label{dlnVdphidN_approx}
\ee
Combining Eqs.~(\ref{dphidN_approx}-\ref{dlnVdphidN_approx}), and that $\mpl ^2d^2\ln V/d\phi^2 = - d\sqrt{2\parslope}/ d\phi$, we obtain
\be
\parrun \approx -  \frac{C\mpl^2}{2}\frac{d^2\ln V}{d\phi^2} = - \frac{C}{4}\frac{d^2\ln V}{d\psi^2}  \ ,
\ee
where 
\be
C\equiv \frac{8}{3}\left(1-\frac{\ln{(1+\sqrt{2})}}{\sqrt{2}}\right) \approx  1.005 \ .
\ee
For phantom models $C$ is negative.

\begin{figure}
\centering
\includegraphics[width=\narrowfigurewidth]{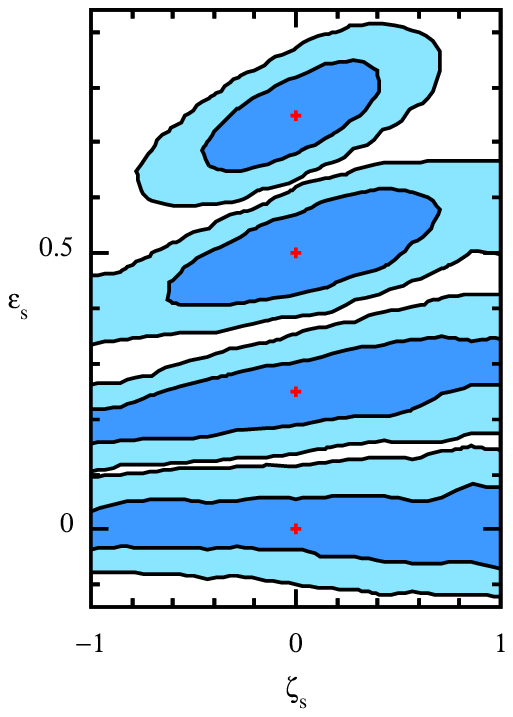}
\caption{The 68.3\% CL (inner contours) and 95.4\% (outer contours) CL constraints on $\parslope$ and $\parrun$, using forecasted CMB, WL, BAO and SN data. The thawing prior ($\partrack=0$) has been used to break the degeneracy between $\parslope$ and $\partrack$. The four input fiducial models labeled with red points have $\parrun=0$ and, from bottom to top, $\parslope=0$, $0.25$, $0.5$, and $0.75$, respectively. Only for large gradients can $\parrun$ be measured and a reasonable stab at potential reconstruction be made} 
\label{figeps0.25}
\end{figure}

We have shown in \S~\ref{sec:forecast} that, for the fiducial $\Lambda$CDM model (with $\parslope=\partrack=0$), the running parameter $\parrun$ cannot be measured. For a nearly-flat potential,  the field momentum is constrained to be very small, hence so is the change $\delta\phi$, so the second order derivative of $\ln V$ is not probed. To demonstrate what it takes to probe $\parrun$,  we ran simulations of fiducial models with varying $\parslope$ ($0$, $0.25$, $0.5$, and $0.75$). The resulting forecasts for the constraints on $\parslope$ and $\parrun$, are shown in Figure~\ref{figeps0.25}. We conclude that unless the true model has a large $\parslope \gtrsim 0.5$, which is  disfavored by current observations at nearly the $2\sigma$ level, we will not be able to measure $\parrun$. Thus if the second derivative of $\ln V$ is of order unity or less over the observable range, as it has been engineered to be in  most quintessence models, its actual value will not be measurable. However, if $|\mpl^2d^2\ln V/d\phi^2| \gg 1$, the field would be oscillating. Though  in their own right, oscillatory quintessence models are not considered in this work.

\subsection{Field Momentum and the Tracking Parameter $\partrack$}

\begin{figure}
\centering
\includegraphics[width=\narrowfigurewidth]{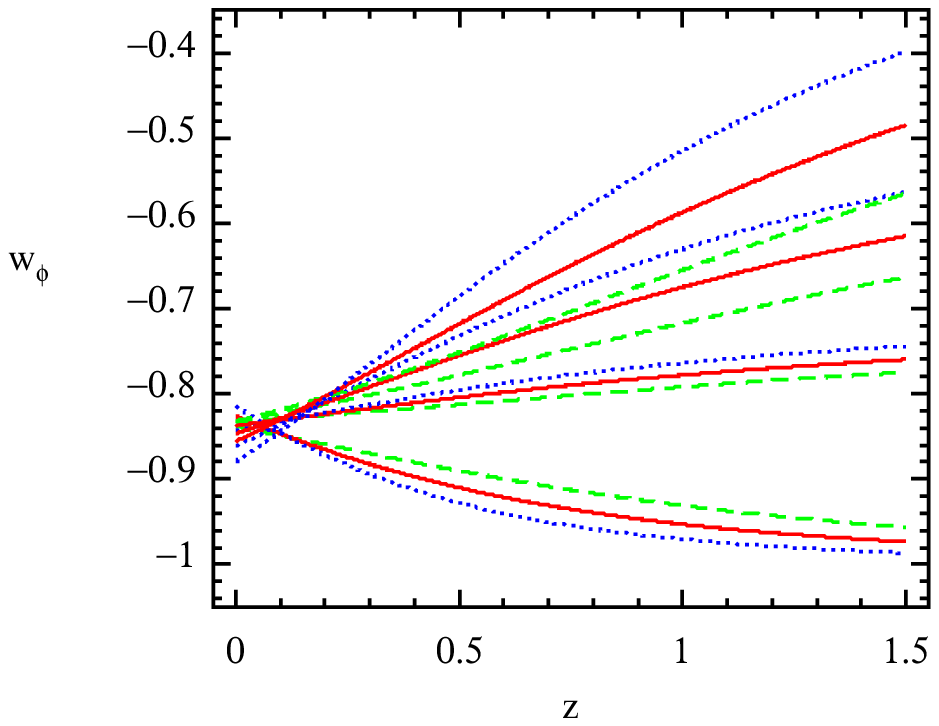}
\caption{The dependence of $w_{\de}(z)$ on $\partrack$ and $\parrun$. We have fixed $\omm=0.27$ and $\parslope=0.5$. The red solid lines correspond to $\parrun=0$, green dashed lines $\parrun=-0.5$, and blue dotted lines $\parrun=0.5$. For each fixed $\parrun$, the lines from bottom to top correspond to $\partrack=0$, $0.3$, $0.6$, and $0.9$. }
\label{figwdep}
\end{figure}

The late universe acceleration requires the field to be in a slow-roll or at most a moderate-roll at low redshift. In tracking models, the high redshift $\epsilon_{\de}$ constancy implies the kinetic energy density follows the potential energy density of the scalar field: $\Pi_\phi^2/2V \rightarrow \partrack/(3-\partrack)$. The field could be fast-rolling in the early universe, with large $\partrack$ and a steep potential, and indeed is the case for many tracking models. If the potential is very flat, field momenta fall as $a^{-3}$, The rate of Hubble damping in the flat potential limit is $\dot\phi \propto a^{-3}$. However, as proposed in most tracking models, the potential is steep at high redshift, and gradually turns flat at low redshift. The actual damping rate itself is related to the field momentum, which again relies on how steep potential is at high redshift. The complicated self-regulated damping behavior of field momentum is encoded in the tracking parameter $\partrack$ in Eq.~\refeq[3parameps] (see Figure~\ref{figwdep} where the damping of field momentum is visualized in $w(z)$ space). At high redshift the approximation $\dot\phi\sim a^{-3}$ completely fails, but at low redshift the damping rate asymptotically approaches $\dot\phi \sim a^{-3}$. 

Since dark energy is subdominant at high redshift, the observational probes there are not very constraining. In the 3-parameter approximation, instead of an asymptotic $\epsilon_{\de \infty }$, 
we could use a moderate redshift  pivot $z_{\rm pivot}\sim 1$, with variables $\varepsilon_{\phi \rm pivot}$, $\parslope$, and $\parrun$, and the asymptotic regime now a controlled extrapolation.
Indeed, using the method of \S~\ref{sec:w0winfty} we could define everything in terms of  pivots of $\epsilon_{\de}$; for example, at $z=0$ as we used in \S~\ref{sec:w0winfty}, and with a third about half way in between 0 and 1, such as at $\aeq$, which is about 0.7, so $z_{\rm eq}\sim 0.4$. Since all would be functions of our original 3 parameters and $\aeq$ through the specific evaluations of $F$ and $F_2$ at the pivots, the formula does not change, but the measure (prior on the parameters) would be different, namely to one uniform in the three $\varepsilon_{\phi \rm pivot}$s. 

Whatever the specific parameters chosen, in all cases the lack of high redshift constraining power means that our reconstructed $w_{\de}(a)$ trajectories such as in Fig.~\ref{figtrajs} should be regarded as low-redshift observational constraints extrapolated through theory-imposed priors on the smooth change in quintessence paths at high-redshift. The tracking assumption that we have used in this paper provides one such extrapolation, but others may be considered. For these, we need to define a prior on $\dot\phi_{\rm pivot}$ and information on the damping rate of $\dot\phi$, and this could have an impact on the marginalized posterior likelihood of $\parslope$. We have presented an example of this, showing that whereas the flat prior $0\le \vert \partrack \vert <\epsilon_m$ gives $\parslope = 0.00^{+0.18}_{-0.17}$, the $\partrack=0$ ``thawing prior''  gives $\parslope=-0.00^{+0.27}_{-0.29}$. We note though that because DE is so subdominant at higher redshift,  observables such as the Hubble parameter $H(z)$, the luminosity distance $d_L(z)$, and the linear perturbation growth rate $D(z)$ are insensitive to the high-$z$ extrapolation of $w_{\de}(z)$, as we showed in Figure~\ref{figtrajs},  Thus, even for the SUGRA model shown in the middle panel  of Figure~\ref{wfit_tracking} with a time-varying attractor which is not fit as well at high $z$, what the data says about it is confined to low $z$, and hence the results obtained with our formula apply to such models as well; in particular the late time potential steepness must be controlled according to the $\parslope$ constraint for them to be viable.

\subsection{Using Our $w_\phi$ Parametrization}

Suppose that one has in mind a specific quintessence potential $V(\phi)$. How can our constraints, in particular on $\epsilon_s$, be used to test viability? The first thing is to see what $\epsilon_V$ looks like at a function of $\phi$. One does not know $\phi_{\rm eq}$ of course, but it had better lie in the range for $\epsilon_s$ allowed by the data. The model would be ruled out if no $\phi$ has $\epsilon_V$ penetrating the allowed region. Otherwise, the field equation of motion has to be solved to get $\phi_{\rm eq}$, and hence $\epsilon_s$ from its definition. Such a calculation is inevitable because quintessence models do not solve the fine-tuning problem.  For tracking models, the low-redshift dynamics is usually designed (i.e., are fine-tuned) to deviate from an attractor. For thawing models, the initial value of $\phi_\infty$ (to which  $\phi$ is frozen at high redshift) needs to be fine-tuned as well. 

Although it seems unlikely that the community will abandon the DETF-sanctioned $w_0$-$w_a$ linear $a$ model and its Figure of Merit as a way to state results for current and future DE experiments, our easy-to-use  $\parslope$  constraint with its clear physics-based meaning could be quoted as well. The 2-parameter case goes well beyond linear-$a$ with the same number of parameters. Because we learn little about the third parameter one might think the 3-parameter case has one parameter too many to bother with, but it is not really redundant since it helps to accurately pave the $w_{\de}$ paths.  To facilitate general-purpose use of this parametrization, we wrote a fortran90 module that calculates  $w_\phi(a\vert \parslope, \partrack, \parrun , \omm)$ (http://www.cita.utoronto.ca/$\sim$zqhuang/work/wphi.f90). The f90 code is well-wrapped so that implementing it in COSMOMC is similar to what has to be done to install the $w_0$-$w_a$ parameterization to deliver more physically meaningful MCMC results. 

\acknowledgments{We are grateful for useful discussions with Patrick MacDonald, Andrei Frolov, and Dmitry Pogosyan.}

\appendix

\section{Comparison with Other Parameterizations}

 The  constant $w$ model and the  linear $w$ models, $w=w_0+w_a(1-a)$ or $w=w_0+w_1z$ with a cutoff \cite{Upadhye2005, Linder2005}, are the simplest $w_{\rm de}$ parameterizations widely used in the literature. An advantage is that they can fit many dark energy models, including those beyond scalar field models, at low redshifts. A disadvantage is that they fail at $z\gtrsim 1$ for most physical models, and in this high-precision cosmology era we cannot ignore adequate inclusion of the $z\sim 1$ information. For this reason there are various three-parameter or four-parameter approximations proposed to improve $w$  at higher redshifts. Some examples are simple extensions of $w_0$-$w_a$: 

\begin{enumerate}
\item[]{
An expansion quadratic in $(1-a)$ with a $w_b$ parameter added to $w_0,w_a$ \cite{Seljak2005,Upadhye2005,Linder2005}, 
\begin{equation}
w(a)=w_0+w_a(1-a)+w_b(1-a)^2\ . 
\end{equation}

}

\item[]{
Replacement of $(1-a)$ by $(1-a^b)$, with $b$ a parameter added to $w_0,w_a$  \cite{Linder2005}, 
\begin{equation}
w(a)=w_0+w_a(1-a^b)\ .
\end{equation}
}

Four parameter models include:
\item[]{
\begin{equation}
w(a)=w_0w_1\frac{a^p+a_s^p}{w_1a^p+w_0a_s^p},
\end{equation}
where $w_0,w_1,a_s,p$ are the parameters \cite{Hannestad2004};
}

\item[]{
\begin{equation}
w(a)=w_0+(w_m-w_0)\frac{1+e^{a_c/\Delta}}{1+e^{(a_c-a)/\Delta}}\frac{1-e^{(1-a)/\Delta}}{1-e^{1/\Delta}},
\end{equation}
where $w_0,w_m,a_c,\Delta$ are constants \cite{Corasaniti2002};
}

\item[]{
\begin{equation}
w(a)=w_f+\frac{\Delta w}{1+(a/a_t)^{1/\tau}},
\end{equation}
where $w_f,\Delta w, a_t,\tau$ are the parameters \cite{Linder2005}.
}

\end{enumerate}

Many of these four-parameter models describe a $w_{\de}$-transition characterized by an initial $w$, today's $w$, a transition redshift, and a duration of transition. For the tracking models they are meant to describe, such a  phenomenology introduces three parameters to describe essentially one degree of freedom (the tracking attractor). Instead our parametrization consistently solves the tracking $w_{\de}(a)$. 

\citet{Chiba2009b} has done similar work for the negative power-law tracking models, though did not model the low-redshift slow-roll regime, making his parameterization much less useful than ours for comparing with data.

Slow-roll quintessence models have been studied by, e.g.,  \citet{Crittenden2007,Scherrer2008,Chiba2009a}, but our work makes significant improvements. Firstly, we have chosen a more optimal pivot $a=\aeq$ to expand $\ln V(\phi)$, whereas the other works expand $V(\phi)$ at either $a\ll 1$ or $a=1$. As a result, we approximate $w_{\de}$ at the sub-percent level in the slow-roll regime $|1+w|<0.2$ compared with the other approximations failing at $|1+w_{\de}|>0.1$. Secondly,  with the field momentum from the tracking regime incorporated, our three-parameter $w_\phi(a)$ ansatz can also fit the more extreme moderate-roll and tracking models with an accuracy of a few percent. Since current data  allow a relatively large region of parameter space ($|1+w_{\de}|\lesssim 0.2$ at 99.7\% CL), a slow-roll prior with $|1+w_{\de} |\ll 1$ should not be imposed as other papers have done. Thirdly, we have extended the parameterization to cover phantom models, though that is arguably not a virtue.
 
\subsection{$\rho_{\rm de}(z)$ and $H(z)$ reconstruction}

One has to integrate $w_{\de}(z)$ once to get the the dark energy density $\rho_{\de}(z)$ and the Hubble parameter $H(z)$.  Observational data are often directly related to either $H(z)$ or $\rho_{\rm de}(z)$, and not sensitive to $w_{\rm de}(z)$ variations. Thus one may prefer to directly parameterize $\rho_{\rm de}(z)$ \cite{Wang2004,Alam2004} or $H(z)$ \cite{Sahni2003,Alam2004,Alam2007}  on phenomenological grounds. A semi-blind expansion of $H(z)$ or $\rho_{\rm de}(z)$, e.g., in polynomials or in bands,  differs from one in $w_{\rm de}(z)$, since the prior measures on the coefficients are radically altered. By contrast, a  model such as ours for $w_{\rm de}(z; \parslope, \partrack, \parrun)$ characterized by physical parameters is also a model for  $H(z; \parslope, \partrack, \parrun)$ or $\rho_{\rm de}(z; \parslope, \partrack, \parrun)$, obtained by integration. 

Where the freedom does lie is in the prior measures imposed on the parameters, a  few examples of which were discussed in Section~\ref{sec:constraint}.

\subsection{$V(\phi)$ reconstruction}

For thawing models, our parameters $\parslope$ and $\parrun$ are based on a local expansion of $\ln V(\phi)$ at low redshift. One should obtain similar results by directly reconstructing the local $V(\phi)$.
An attempt at local $V(\phi)$ reconstruction was done in \citet{Sahlen2005}, where a polynomial expansion was  used, 
\begin{equation}
V(\phi)=V_0+V_1\phi+V_2\phi^2+...\ ,
\end{equation}
with $\phi$ chosen to be zero at present. A result highlighted in that paper is that there is a strong degeneracy between $V_1$ and $\dot\phi_0$. We shall see that this degeneracy is actually a result of the prior, and not something that observational data is telling us, as we now show.

For simplicity we assume that $\omm$ and $h$ are known, hence the parameter $V_0$ is hence fixed. Although $V_1$ is defined at the pivot $a=1$, it is approximately proportional to $\sqrt{\parslope}$ since $\epsilon_V$ varies slowly. Because $\dot\phi_0$ is a function of $1+w|_{z=0}$, the degeneracy between $V_1$ and $\dot\phi_0$ is roughly the degeneracy between $\parslope$ and $\epsilon_{\de 0}$, an expression of the low-redshift dynamics being mainly dependent upon the slope of $\ln V$. This is explicitly shown in Figure~\ref{figwdep}. 

\citet{Huterer2007} tried to do $V(\phi)$ reconstruction using what was then recent observational data. They assumed zero field momentum at high redshift ($z_{\rm start}=3$ is used), which is essentially the thawing prior, and used RG flow parameters $\epsilon_V$, $\eta_V$, etc. expanded about the initial redshift $z_{\rm start}=3$. The constraint they find on the parameter $\epsilon_V|_{z=3}$ is very weak. From our work, this is understandable because the high-redshift dynamics are determined by the tracking parameter rather than by $\varepsilon_V$. They also calculated the posterior probability of $\varepsilon_V$ at redshift zero, which is much better constrained. Their result is consistent with what we have obtained for $\parslope$. Also the large uncertainty in $\eta_V|_{z=0}$ they find is similar to our result for $\parrun$.

Recently \citet{Fernandez2008} have proposed Chebyshev expansions of $V(z)$ or $w_{\de}(z)$, a method we have applied to treat early universe inflation \cite{scanning}, but as we have discussed in the introduction there is de facto much less information in the $\sim$ one e-folding in $a$ that DE probes cover than the $\sim$ 10 e-foldings in $a$ that CMB+LSS power spectrum analyses cover.

\end{document}